\documentclass[a4paper,10pt]{article}

 \usepackage[cp1251]{inputenc}
 \usepackage{cite}

 \usepackage{graphicx}

\usepackage{cite,amsmath,amsfonts,amsthm,fullpage}
\usepackage{youngtab}

\newcommand{\bpow}{\mathbf{p}}

\newcommand{\tig}{\tilde{\Gamma}}

\newcommand{\gF}{\mathfrak{F}}
\newcommand{\gC}{\mathfrak{C}}

\newcommand{\fa}{\textsc{f}}
\newcommand{\e}{\textsc{e}}
\newcommand{\V}{\textsc{v}}

\newcommand{\mc}{\textsc{m}}

\theoremstyle{plain}

\newtheorem{Lemma}{Lemma}
\newtheorem{Proposition}{Proposition}
\newtheorem{Corollary}{Corollary}
\newtheorem{Remark}{Remark}

\theoremstyle{remark}

\def\g{\Gamma}

\def\bp{\begin{Proposition}}
\def\ep{\end{Proposition}}
\def\bc{\begin{Corollary}}
\def\ec{\end{Corollary}}
\def\bl{\begin{Lemma}}
\def\el{\end{Lemma}}
\def\be{\begin{equation}}
\def\ee{\end{equation}}
\def\br{\begin{Remark}\rm\small}
\def\er{\end{Remark}}
\def\brs{\begin{remarks}.\\ \rm\
\begin{enumerate}}
\def\ers{\end{enumerate}\end{remarks}}
\def\bea{\begin{eqnarray}}
\def\eea{\end{eqnarray}}

\def\tr{\mathrm {tr}}
\def\det{\mathrm {det}}

\def\diag{\mathrm {diag}}

\def\diag{\mathrm {diag}}

\def\&{&{\hskip -20pt}}

\usepackage[usenames,dvipsnames]{color}
\usepackage{ulem}

\begin{document}

\author{ Sergey M. Natanzon\thanks{National Research University Higher School of Economics, Moscow, Russia; 
Institute for Theoretical and Experimental Physics, Moscow, Russia;
email: natanzons@mail.ru} \and Aleksandr Yu.
Orlov\thanks{Institute of Oceanology, Nahimovskii Prospekt 36,
Moscow 117997, Russia, and Institute for Theoretical and Experimental Physics, Moscow, Russia, 
email: orlovs@ocean.ru
}}
\title{Hurwitz numbers from Feynman diagrams}

\maketitle

\begin{abstract}

\noindent 

Our goal is to construct a generating function for Hurwitz numbers of the most general type: with an arbitrary 
base surface and arbitrary branching profiles.
We consider a matrix model constructed according to a graph on an orientable connected surface 
$ \Sigma $ without boundary. If we call $ \Sigma $ a sky, then the verticies are stars, and the graph is 
a children's drawing of a constellation (dessins d'enfants). We consider stars as small circles. 
We put matrices on the segments of the circles (we call them source matrices), their product 
defines the monodromy of a given star; the monodromy spectrum we refer as the spectrum of this star. $\Sigma$ 
consists of glued charts, each chart corresponds to  a combination of random matrices and source matrices; 
Wick's  pairing is responsible for gluing the surface from the set of charts. Additional gluing of the Moebius 
stripes into $ \Sigma $ corresponds to introducing a special tau function into the measure of integration. 
The matrix integral can be evaluated as a series in terms of the spectrum of stars, and the coefficients of 
the series are Hurwitz numbers, which count the coverings of the surface $ \Sigma $ (and also of its Klein 
extension by pasting Moebius stripes) with any given set of branching profiles at the vertices of the graph. 
In this paper, the emphasis is on the combinatorial description of the matrix integral. 
Hurwitz number is equal to the number  Feynman diagrams of a certain type, divided
by the order of the automorphism group of the graph.

\end{abstract}

\bigskip


\textbf{2010 Mathematic Subject Classification:} 05A15, 14N10, 17B80, 35Q51, 35Q53, 35Q55, 37K20, 37K30,

\bigskip

\bigskip

\section{Introduction}

This work is an expanded version of the report of one of the authors (A.O.) at a conference 
Workshop on Classical and Quantum Integrable Systems in Euler Institute, 22.07.2019-26.07.2019
in St. Petersburg.
In the report, several points were briefly noted:

(1) matrix integrals were presented, constructed according to a 
graph at the vertices of which are placed source matrices that play the role of coupling constants in the matrix model.
The answers for such integrals were expressed as series in Schur functions. The same series is a generating function 
for general Hurwitz numbers, that is, in the case when the covered (base) surface is an orientable (or possibly 
non-orientable) connected surface without a boundary with any Euler characteristic 

(2), this integral was calculated 
by the decomposition method by character. 

(3) tau functions were used as an integrand

(4) the connection between the problem of finding the Hurwitz numbers and the two-dimensional Yang-Mills theory was 
shown
(5) the integral was analyzed by the Feynman diagram method 

(6) the combinatorial aspects of these problems were considered. 

The text with the development of paragraphs (2) - (3) - (4) was published in \cite{NO-2019} and \cite{NO2020}
 This work partially repeats these works, but with 
different accents, and item (6) is also considered in more detail.

We note that quite a rich literature is noted in the connection of matrix integrals with Hurwitz numbers. 
However, various special cases of generating functions were considered everywhere, and not for the Hurwitz numbers 
themselves, but for some linear combinations. Our approach is universal and allows us to find general Hurwitz numbers.

Let us write down the references which can be related to the topic 
(we apologise unnamed authors of important works: this list is obviously incomplete).

Hurwitz numbers: \cite{H},\cite{Fr},\cite{FS}, \cite{M1}, \cite{M2},\cite{GARETH.A.JONES},\cite{N}
\cite{AN1},\cite{AN2008},\cite{AN3},\cite{AN3},\cite{MM1},\cite{MM3}.

Hurwitz numbers in the context of string theory, a key observation was made by Dijgraaf in
 \cite{Dijkgraaf},\cite{{D2}}.
It was followed by works \cite{Okounkov-2000},\cite{Okounkov-Pand-2006},\cite{ELSV},\cite{KazarianLando},
\cite{MM2},\cite{MM5} and many others.

Combinatorial aspects related to the covering problem: \cite{GJ}, \cite{Goulden-Jackson-2008},
\cite{Goulden-Paquet-Novak},\cite{GGPN}.

Klein surfaces in the context of our topics: \cite{CGN}, \cite{KLN}.

Matrix models and orientable surfaces: a key observation was made by  t'Hooft in \cite{t'Hooft}.
Then, a lot of important applications and developements of the topic of matrix integrals
 is contained in  important papers \cite{Itzykson-Zuber},
\cite{BrezinKazakov},\cite{GrossMigdal},
\cite{Kazakov},\cite{Kazakov2},\cite{Kazakov3},\cite{Kazakov-SolvMM},\cite{KazakovKostovNekrasov},
\cite{Kazakov-ZinnJ},\cite{NO2020}.

Ginibre enesembles and independent Ginibre ensembles: \cite{FyodorovSommers},\cite{Ak1},\cite{Ak2},\cite{AkStrahov},
\cite{S1},\cite{S2},\cite{Chekhov-2014}.

Matrix models and Hurwitz numbers: \cite{Amburg} ,\cite{MelloKochRamgoolam}, \cite{Alexandrov}, \cite{Zog}, \cite{KZ}, 
\cite{NO-2014},\cite{ChekhovAmbjorn},\cite{Chekhov-2014},\cite{NO-LMP},
\cite{GouldenNovak},\cite{O-TMP-2017},\cite{O-2017},\cite{Orlov-chord}, \cite{DimaVasiliev}.

Integrable systems and Hurwitz numbers: key works - \cite{Okounkov-2000},\cite{Okounkov-Pand-2006},
\cite{Goulden-Jackson-2008}. Further a lot of work was done:
. \cite{AMMN-2011},\cite{Takasaki-Hurwitz},\cite{AMMN-2014},\cite{GayPakettHarnad1},   \cite{Dubr},\cite{NO-2014},
\cite{HO-2014}, \cite{GayPakettHarnad2},\cite{NatanzonZabrodin},
\cite{NO-2019}. 
Overviews: \cite{MM4}, \cite{AMMN-2014}, \cite{Uspehi-KazarianLando}, \cite{HarnadOverview}.

Topological theories and Hurwitz numbers: \cite{Dijkgraaf},\cite{{D2}}, \cite{AN1},\cite{AN2008},
\cite{AN3},\cite{GN},\cite{LN}, \cite{Nat}.

The main goal of this paper is to provide a combinatorial description of the integrals (\ref{MM1-all'})
- formulas (\ref{Hurwitz-combinatorial}) and (\ref{MM1-all''}), which describes the Feynman diagrams of the integral
(\ref{MM1-all}) and similar integrals: ()(\ref{MM1-all}),(\ref{full-generating function}). The expansion of the integral (\ref{MM1-all}) is carried out over the insertion matrices (
source matrices), more precisely, over  spectral functions of their products (over `` spectrum of stars ''). 
The Feynman graph  of the lowest order is the basement of the matrix model. The matrix model is built according to its
first Feynman graph.
The combinatorial meaning of the lowest graph is given by a relation  (\ref{combinatorial-graph})) in 
the symmetric group $ S_ {2n} $, where $ n $ is the number of ribbon edges in the graph ($ 2n $ is the number of 
matrices of the multi-matrix integral).
This relation is well known as a combinatorial description of maps; see the wonderful book \cite{ZL}. Higher
orders of perturbation theory describe the coverings of the lowerest Feynman graph and are a generating function 
for Hurwitz numbers.

Some other topics are briefly discussed (integration of tau functions and non-orientable coverings). 

In Section \ref{Discussion} titled Discussion, we develope the work \cite{MM1},
which offers a beautiful generalization of the cut-and-join formula (MMN formula):
\be\label{MM3'}
{\cal W}^{\Delta}(\bpow)\cdot s_\mu(\bpow)=\varphi_\mu(\Delta)s_\mu(\bpow),
\ee
which describes the merging of pairs of branch points in the covering problem.
Here $\mu=(\mu_1,\mu_2,\dots)$ and $\Delta=(\Delta_1,\dots,\Delta_\ell )$  are Young diagrams 
($\Delta$ is  the ramification profile of one of branch points;
for simplicity, we consider the case where $|\mu|=|\Delta|$),
$s_\mu$ is the Schur function.
Differential operators $ {\cal W}^{\Delta} (\bpow) $ generalize the operators of "additional symmetries"
\cite{O-1987} in the theory of solitons and commute with each other for different $ \Delta $.
In the work (\cite{MM3}), it was noted that if they are written in the so-called Miwa variables, that is,
in terms of the eigenvalues of the matrix $X$ such that 
$ p_m = \tr \left( X^m \right) $, then
the generalized cut-and-join formula (\ref {MM1}) is written very compactly and beautifully:
\be\label{MM3}
{\cal W}^{\Delta}\cdot s_\mu(X)=\varphi_\mu(\Delta)s_\mu(X)
\ee
where
\be
{\cal W}^{\Delta}=\frac{1}{z_\Delta}\tr\left( D^{\Delta_1} \right)\cdots \tr\left( D^{\Delta_\ell}\right),
\ee
and the factor $z_\Delta$ is given by (\ref{z_Delta}), $D$ is
\be
D_{a,b}=\sum_{c=1}^N X_{a,c}\frac{\partial}{\partial X_{b,c}}
\ee
As G.I.Olshansky pointed out to us, this type of formula appeared in the works of Perelomov and Popov
\cite {PerelomovPopov1}, \cite {PerelomovPopov2}, \cite {PerelomovPopov3}
and describe the actions of the Casimir operators in the representaion $ \lambda $, see
also \cite{Zhelobenko}, Section 9.

We propose a generalization of this relation, which in our case is constructed
using a child's drawing of a constellation (dessins d'enfants, or a map in terminology \cite{ZL}.
In fact, we are considering a modification in which the vertices are replaced by small disks - "stars"). This topic will be
be studied in more detail in the next article. 
Here we restrict ourselves only to a reference to important  beautiful works
\cite{Olshanski-19},\cite{Olshanski-199},\cite{Okounkov-1},\cite{Okounkov-19},\cite{Okounkov-199}.

In the Appendix
some review material on Hurwitz numbers from the literature and from previous works of the authors is given.

\section{Feynman integrals for a model of complex matrices related to a ribbon graph with inflated vertices
\label{integral}}

Consider a ribbon (aka fat) connected graph $ \Gamma $ on an orientable connected surface $ \Sigma $ without boundary with Euler
characteristic $ \e $, where all faces are homeomorphic to disks. Then $ \e = \V-n + \fa $, where $ \V  $ is the number
vertices $ n $ - the number of edges (ribbons), $ \fa $ - the number of faces.
We construct the graph $ \tig $, which is obtained from $ \Gamma $ by replacing the vertices with small disks 
(hereinafter, we will call them small disks or, the same, inflated vertices).
Then $ \tig $ has $ \fa + \V $ faces: $ \V $ small disks and $ \fa $ original faces, 
the latter will be called basic faces to distinguish them from small disks.
Graph $\tig$ has $ 2n $ trivalent vertices  (the endpoints of ribbons), each of which has one outgoing ribbon and
two outgoing segments of small disks. There are $3n$ edges: $n$ ribbons and $2n$ segments of small disks.

 \begin{figure}[h]
  \centering
	 \includegraphics[height=4cm]{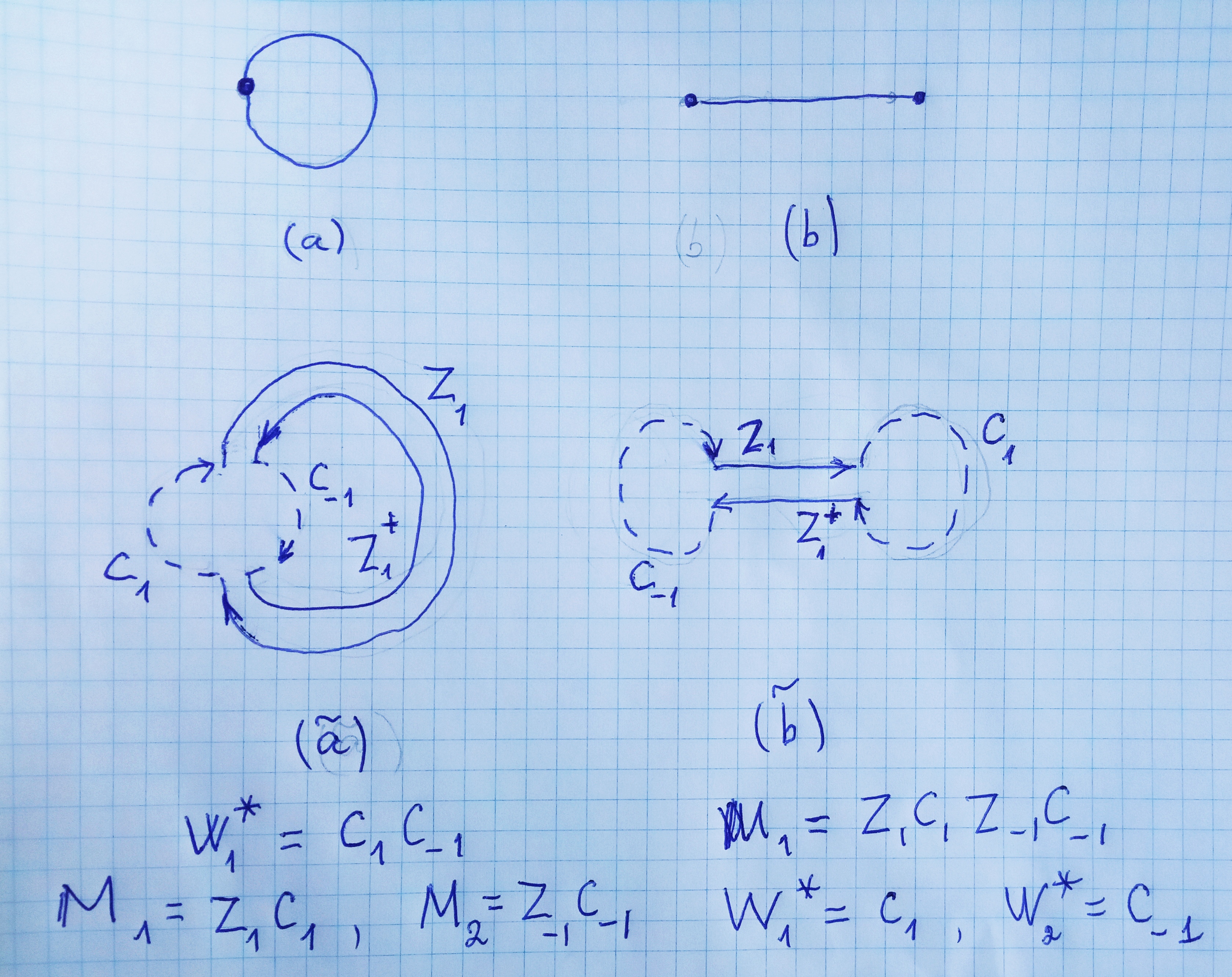}
  \caption{}\label{figure1} Dual graphs $\Gamma$ $(a)$ and $(b)$ in case $n=1$.
  Graphs $\tig$ $(\tilde{a})$ and $(\tilde{b})$ are drawn below them, which relate to graphs $(a)$ and $(b)$, respectively  
 \end{figure}%

 \begin{figure}[h]
  \centering
	 \includegraphics[height=3cm]{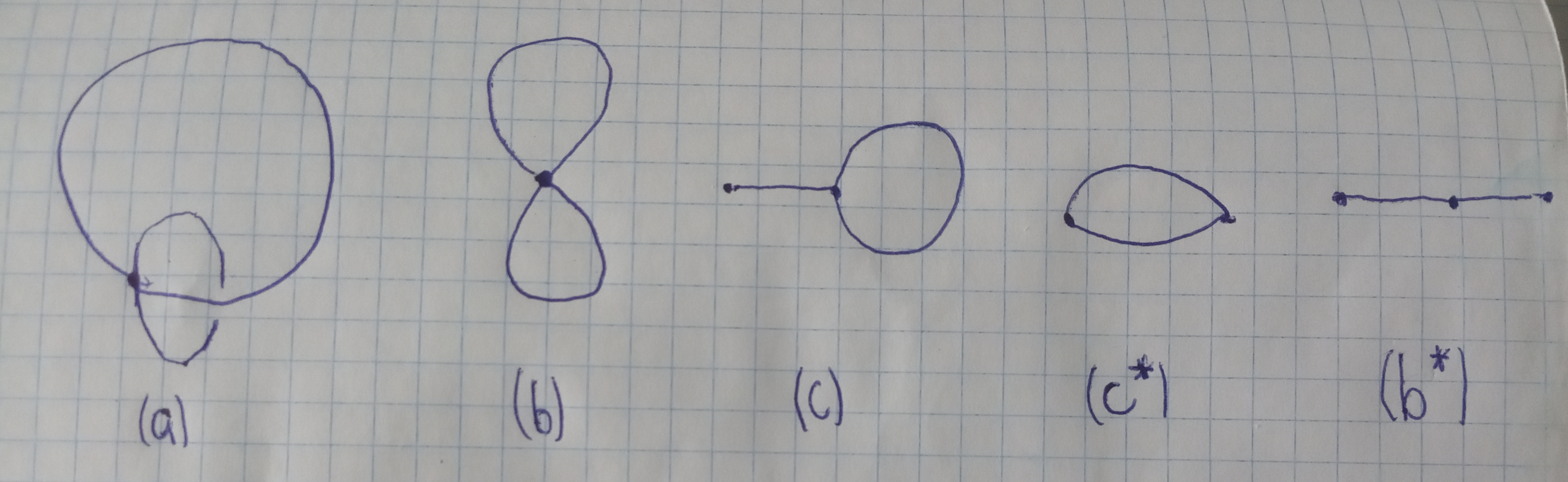}
  \caption{}\label{figure2} All graphs $\Gamma$ with $n=2$. In the case (a), $\Sigma\subset \mathbb{T}^2$, for the cases
  (b)-(e), $\Sigma=\mathbb{S}^2$.
 \end{figure}%
 
 \begin{figure}[h]
  \centering
	 \includegraphics[height=3cm]{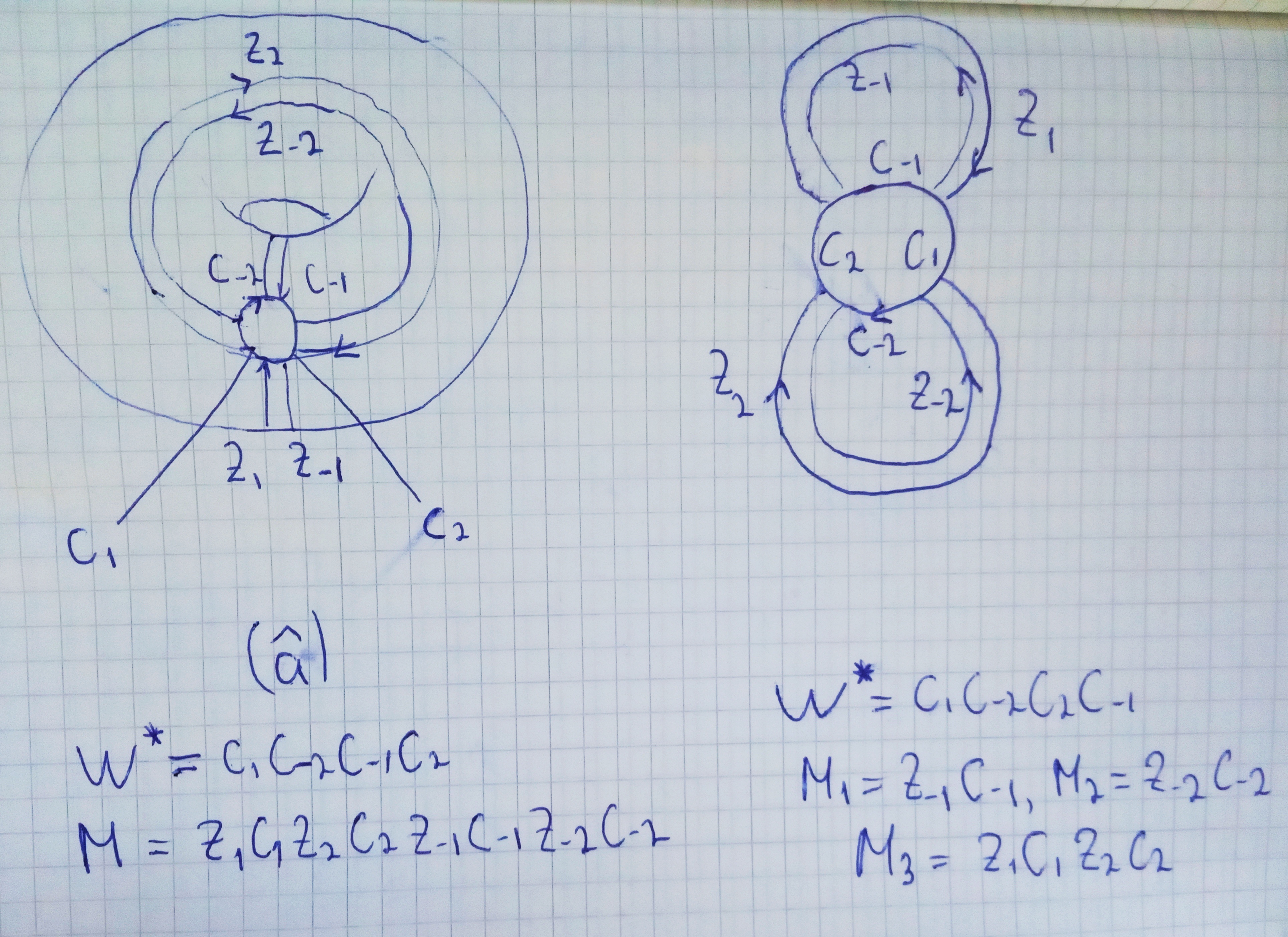}
  \caption{}\label{figure4} Graphs $\tilde{\Gamma}$ for the cases (a) and (b) in fig.\ref{figure2}
 \end{figure}%

We will also assign a positive (counterclockwise) orientation to the sides of the basic faces, then each side of 
each edge of the ribbon will be a solid arrow, and the ribbon itself will be a pair of oppositely directed solid 
arrows. Each solid arrow of the ribbon continues with a dotted arrow - a segment of the boundary of a small disk.

As we see, the boundary  of each small disk consists of a sequence of dashed arrows, each of which is directed 
negatively (clockwise), if you go around the center of the small disk.

The boundary of the basic faces of $\tig$ consists of solid and dotted arrows, which alternate one after another and are 
directed positively if you go around the 'capital' of the basic face.
The 'capital' will be the selected point inside the basic face of $\tig$.

Now we assign to each of the $ 2n $ trivalent vertices $ \tig $ a number from $ 1 $ to $ N $ and a pair of these numbers 
to each arrow according to the end points.

We number the ribbons with numbers from $ 1 $ to $ n $ and
we number the solid arrows with numbers from $ 1 $ to $ 2n $, so that the arrows belonging to the ribbon $ i $ have
opposite signs: $ i $ and $ -i $ (choosing which arrow is numbered $ i $ and which $ -i $ is not important).
We number each dotted arrow with the same number as the solid arrow, which abuts against the beginning of the 
dotted one, see the figure.

Finally, with a solid arrow with such a set, we associate a certain number
$ (Z_i)_{a, b} \in \mathbb {C} $ ($ i = \pm 1, \dots, \pm n $),
and with the dashed arrow with this set we associate the number $ (C_i)_{a, b} $ ($ i = \pm 1, \dots, \pm n $).

We consider these complex numbers to be matrix elements of complex matrices
$$
Z_{\pm 1}, \dots, Z_{\pm n}, \quad C_{\pm 1}, \dots, C_{\pm n},
$$
with the condition
$$
Z_{-i} = Z^\dag_i, \quad i = 1, \dots, n
$$
The matrices $\{C_i\}$ we call the source matrices which play the role of coupling constants in the matrix models 
below.

A graph $\tig$ equipped with arrows and numbers in this way will be called the equipped graph $\tig(\{Z_i,C_i\})$.

We will number small disks with numbers from $ 1 $ to $ V $.
Let us call small disks (inflated vertices) stars.
For each star (small disk) of the graph $ \tilde{\Gamma} (\{Z_i, C_i \}) $
we introduce a monodromy of the star: the product of such matrices from the set $ \{C_i \} $ that are assigned to the 
dashed arrows along the boundaries
of this disc in the order indicated by the sequence of these arrows, following one after the other clockwise.
\be\label{star-monodromy}
W_i^* =C_{i_1}\cdots C_{i_k},\quad i=1,\dots,\V
\ee
where matrices $C_{i_1},\dots ,C_{i_k}$ correspond to the $k$ dashed arrows attached to each other sequentially 
clockwise around the star $i$, from which $k$ ribbons come out.

We determine the monodromies up to a cyclic permutation of matrices. 

In Appendix \ref{Dual monodromies}, a purely algebraic method is written in $n$ steps to obtain a 
set $\{W^*\}$ from the set $\{M\}$ and back.
\br
We make a remark about the geometric picture associated with the above.
If the matrix elements are represented by arrows, then the matrix construction can be represented as a chain
from the arrows assigned to each other with the summation of all numbers from the interval $ [1, N] $ assigned to 
the vertices. The trace of the matrix product is closed chains of arrows, that is, polygons. Thus, the trace of 
monodromy is in the coorespondence with the piece of a two-dimensional surface homeomorphic to a disk - with a polygon.
And you can glue surfaces from polygons.
For the first time, a connection between surfaces glued from polygons and matrix integrals was discovered by t'Hoft
\cite{t'Hooft} and was actively used in works on two-dimensional quantum gravity in \cite{BrezinKazakov} and
in \cite{GrossMigdal}.
\er

Actually
we need only the spectrum of the monodromies of stars - the ``spectrum of the stars'':
\be\label{spectrum}
{\rm Spect}\,W^*_i =\left(w^*_{i,1},\dots, w^*_{i,N}  \right),\quad i=1,\dots,\V
\ee

Let's also label the basic faces with numbers from $1$ to $\fa$, and
we define the monodromy of the basic face as the product of the matrices that correspond to the arrows when going 
around the 
boundary of the face in the positive direction:
\be\label{face-monodromy}
M_i =Z_{i_1}C_{i_1}\cdots Z_{i_m}C_{i_m},\quad i=1,\dots,\fa
\ee
where matrices $Z_{i_1}C_{i_1},\dots ,Z_{i_m}C_{i_m}$ correspond to the $m$ pairs of solid-dashed arrows attached to 
each other sequentially counterclockwise around the capital $i$, along the boundary which consists of $2m$
arrows.

This monodromy can also be determined up to a cyclic permutation and we need only the spectrum of this matrix:
\begin{equation}\label{Spect-face-monodromy}
{\rm Spect}\,M_i =\left(m_{i,1},\dots, m_{i,N}  \right),\quad i=1,\dots,\fa
\end{equation}

The equipped graph $\tig(\{Z_i,C_i\})$ is the Feynman graph of the lowest order of the following matrix model
\be\label{MM1}
\int \left(\prod_{i=1}^\fa  e^{\tr M_i} \right) d\Omega(Z_1,\dots,Z_n)
\ee
where
$$
d\Omega(Z_1,\dots,Z_n)=c_{N}^n\prod_{i=1}^n \prod_{a.b=1}^N e^{-N|(Z_i)_{a,b}|^2}d^2(Z_i)_{a,b},
$$
with normalization
$$
\int d\Omega(Z_1,\dots,Z_n)=1.
$$
The set $\{Z\}$ together with the measure $d\Omega(Z_1,\dots,Z_n)$ is known as $n$ independent complex
Ginibre ensembles. \footnote{ In particular, such ensembles are used in the theory of quantum chaos and information
transfer.
In \cite{NO-2019}, we showed that they are also suitable for describing the two-dimensional Yang-Mills theory,
which is close to the description in \cite{Rusakov}, \cite{Witten}.  }

In our problem, the number $n=1,2,\dots$ is considered given and instead of $d\Omega(Z_1,\dots,Z_n)$ 
everywhere we will write $d\Omega$ .

Namely, $\tilde\Gamma$ is the Feynman graph of the integral
\be\label{d=1}
\int \left(\prod_{i=1}^\fa  \tr M_i \right) d\Omega=N^{-n}\prod_{i=1}^{\V} \tr W^*_i \,.
\ee

This equality will be discussed later, and now write the answer for all orders in the expansion of the integral
(\ref{MM1}). 
For $d_1,\dots, d_\fa$ (where we put $d_1=d$), we have
\be\label{MM1-all'}
\int \prod_{i=1}^\fa
\frac{\left(\tr ( M_i )   \right)^{d_i}}{d_i!}
d\Omega
= \delta_{d,
d_1,\dots,d_\fa}  
N^{-nd}
\sum_{{\Delta}^1,\dots , {\Delta}^{\V}\atop |\Delta^1|=\cdots = |\Delta^{\V}|=d} 
H_\Sigma({\Delta}^1,\dots , {\Delta}^{\V})
{C}({\Delta}^1,\dots , {\Delta}^{\V} ),
\ee
where $\delta_{d, d_1,\dots,d_\fa}=1 $, if $d_1=\cdots = d_\fa$ and $\delta_{d, d_1,\dots,d_\fa}=0 $ otherwise.

The important point is that due to the arbitrariness of the source matrices, (\ref{MM1-all'}) is not one, 
but a family of relations. This fact will be used later in Section \ref{Discussion} when we will construct differential operators
related to $\tig$.
Notice that the source matrices come in different combinations on the left and right sides of the equality.

We obtain the following model
\begin{equation}\label{MM1-all}
\int \left(\prod_{i=1}^{\fa}  e^{\tfrac{1}{\hbar} {\tr} M_i} \right) d\Omega=
\sum_{d=0}^\infty (\hbar N)^{-nd}
\sum_{{\Delta}^1,\dots , {\Delta}^{\V}\atop |\Delta^1|=\cdots = |\Delta^{\V}|=d} 
H_\Sigma({\Delta}^1,\dots , {\Delta}^{\V})
{C}({\Delta}^1,\dots , {\Delta}^{\V} ),
\end{equation}
which we treat as a formal series in powers of the parameter $\hbar^{-1}$ and
where 
\begin{equation}
 {C}({\Delta}^1,\dots , {\Delta}^{\V} ) := \prod_{i=1}^{\V}  \prod_{k=1}^\infty
 {\tr}\left( W_i^* \right)^{\Delta^i_k} 
\end{equation}
is a quantity that depends only on the spectrum of stars, and where $ H_\Sigma({\Delta}^1, \dots, {\Delta}^{V}) $ is 
\textit{the Hurwitz number} enumerating coverings of degree $ d $ of an orientable connected surface without boundary
$ \Sigma $ with branch profiles of type $ {\Delta}^1, \dots, {\Delta}^V $ at $ {V} $ points; for precise 
definition of Hurwitz numbers see Appendix. The branching profile of $ \Delta^i $ is a Young diagram
$ \Delta^i = \left (\Delta^i_1, \Delta^i_2, \dots \right) $, which indicates how the sheets which cover  the 
surface $ \Sigma $ merge. For clarity, the branch points can be considered the centers of stars, although, 
as is known, the Hurwitz numbers of surfaces without boundaries are independent of the location of the branch points.

In case $d=1$ (the first perturbation order) we obtain (\ref{d=1}) since $H\left((1),\dots,(1) \right)$ which 
descibes the covering of $\Sigma$ by itself is equal to $1$.

The formula (\ref{MM1-all}) was proved in \cite{NO2020} geometrically, based on the definition of the Hurwitz number,
 as the weighted number of the ways to glue the covering surface from polygons.

\br \label{on-tau-function}
A generalization of the integral consists in replacing the integrand by the function tau or, in a more general case,
 product of tau functions. (For example, by evaluation of the integral of a product of certain tau
 functions  can be obtained well-known \cite{Rusakov}, \cite{Witten} correlation functions of the two-dimensional gauge
 theory, see \cite{NO2020}).
 Let us dwell on the first option.

The tau function of the multicomponent KP equation \cite{NovikovManakovZakharov} has the form \cite{Sato}:
\begin{equation}
\tau_g(M_1,\dots,M_\fa)=\sum_{\lambda^1,\dots,\lambda^{\fa}} g(\lambda^1,\dots,\lambda^{\fa}) 
s_{\lambda^1}(M_1)\cdots s_{\lambda^{\fa}}(M_{\fa})
\end{equation}
where each $\lambda^i=\left(\lambda^i_1,\lambda^i_2,\dots  \right)$ is a partition and
where
$s_{\lambda^i}(M_i)$ is the Schur function defined as follows \cite{Mac}
$$
s_{\lambda^i}(M_i)=\frac{\det\left[m_{i,k}^{\lambda^i_k-k+N}  \right]_{1\le j,k\le N}}
{\det\left[m_{i,k}^{-k+N}  \right]_{1\le j,k\le N}}
$$
(see (\ref{Spect-face-monodromy})) and where $g(\lambda^1,\dots,\lambda^{\fa}) $ solves certain equation whose form 
we will not specify.
Let us note that the matrix model (\ref{MM1}) is related to the case where
$$
 g(\lambda^1,\dots,\lambda^{\fa})=\prod_{i=1}^{\fa} \frac{\dim\,{\rm\lambda^i}}{d!}
$$
where $\dim\,{\rm\lambda^i}$ is the dimension of the irreducible representation of the permutation 
group $S_d$ labelled by $\lambda^i$.

For even $\fa$ we get \cite{NO-2019}
\begin{equation}\label{MM1-tau}
\int \tau_g(M_1,\dots,M_\fa) d\Omega
\end{equation}
\begin{equation}
=
\sum_{\lambda\atop \ell(\lambda)\le N} N^{-n|\lambda|}
 \left(\frac{\dim\,{\rm\lambda}}{d!}\right)^{ -n}
 g(\lambda,\dots,\lambda)s_\lambda (W^*_1)\cdots s_\lambda (W^*_{\V})
\end{equation}
\begin{equation}
  = \sum_{d=0}^\infty N^{-nd}
\sum_{{\Delta}^1,\dots , {\Delta}^{\V}\atop |\Delta^1|=\cdots = |\Delta^{\V}|=d} 
H_{\Sigma'}(g|{\Delta}^1,\dots , {\Delta}^{\V})
{C}({\Delta}^1,\dots , {\Delta}^{\V} ),
\end{equation}
where
$$
s_{\lambda^i}(W^*_i)=\frac{\det\left[w_{i,k}^{\lambda^i_k-k+N}  \right]_{1\le j,k\le N}}
{\det\left[w_{i,k}^{-k+N}  \right]_{1\le j,k\le N}}
$$
and
\begin{equation}
 H_{\Sigma'}(g|{\Delta}^1,\dots , {\Delta}^{\V})=\sum_{\lambda\atop \ell(\lambda)\le N}
 \left(\frac{\dim\,{\rm\lambda}}{d!}\right)^{\e-\V}
 g(\lambda,\dots,\lambda)\varphi_\lambda (\Delta^1)\cdots \varphi_\lambda (\Delta^{\V})
\end{equation}
can be called the weighted Hurwitz number, similar to the Hurwitz number on the completed cycles 
\cite{Okounkov-Pand-2006}.
Here $ \Sigma'$ can be thought of as $ \Sigma $ with $ \tfrac 12 \fa $ handles inserted.

\er

The non-orientable case corresponds to the matrix model:
\be
\label{MM1-all-klein}
\int 
 \left(\prod_{i=1}^{\fa_1}  e^{\tfrac{1}{\hbar}{\tr} M_i} \right)
\prod_{i=\fa_1+1}^{\fa}
\det\frac{(1+\tfrac{1}{\hbar}M_i)^{\tfrac12}(1-\tfrac{1}{\hbar}M_i)^{-\tfrac12} }
{\left( I_N \otimes I_N - \tfrac{1}{\hbar^2}
M_i\otimes M_i \right)^{\tfrac12}}
  d\Omega
\ee
\be
=\sum_{d=0}^\infty (\hbar N)^{-nd}
\sum_{{\Delta}^1,\dots , {\Delta}^{\V}\atop |\Delta^1|=\cdots = |\Delta^{\V}|=d} 
H_{\tilde\Sigma}({\Delta}^1,\dots , {\Delta}^{\V})
{C}({\Delta}^1,\dots , {\Delta}^{\V} ),
\ee
where $\tilde\Sigma$ is connected non-orientable surface with Euler charactristic $\tilde{\e}=\fa_1 - n +\V$.
We are only interested in the topological structure of surfaces.
Therefore, $ \tilde \Sigma $
 can be interpreted as $ \Sigma $ with glued into it $ \fa - \fa_1 $ Mobius stripes.
This equality (\ref{MM1-all-klein}) formaly can be obtained with the help of (\ref{MM1-all'}),
(\ref{vac-tau-BKP'}) in Appendix \ref{Mobius'} and (\ref{Hurwitz-down}) 
in Appendix \ref{Hurwitz numbers and representation theory}. However there is a geometric interpretation,
see Appendix \ref{Mobius'}.
However, it also has a geometric meaning as a function defined on the so-called orienting
covering of the real projective plane with a hole (that is, defined on the sphere  with involution and two holes  
that twice covers $ \mathbb {RP}^2 $ with a hole, see also Appendix \ref {Mobius'}.

Since we glue the Mobius stripes, why not glue handles.
 We assume that the number $ \fa- \fa_1- \fa_2 $ is even and equal to $ 2h $. Matrix model that will describe
covering the surface $ \Sigma $, in which $ \fa_2- \fa_1=m $ of Moebius stripes and $ h $ are additionally pasted
 handles looks like this:
\be
\label{MM1-all-klein-handle}
\int 
 \left(\prod_{i=1}^{\fa_1}  e^{\tfrac{1}{\hbar}{\tr} M_i} \right)
 \left(
\prod_{i=\fa_1+1}^{\fa_2}{\mathfrak{M}}(M_i) \right)
\left(\prod_{i=\fa_2+2,\fa_2+4,\dots}^{\fa}
{\mathfrak{H}}(M_{i-1},M_{i})\right)
  d\Omega
\ee
\be
=\sum_{d=0}^\infty (\hbar N)^{-nd}
\sum_{{\Delta}^1,\dots , {\Delta}^{\V}\atop |\Delta^1|=\cdots = |\Delta^{\V}|=d} 
H_{\tilde{\tilde\Sigma}}({\Delta}^1,\dots , {\Delta}^{\V})
{C}({\Delta}^1,\dots , {\Delta}^{\V} ),
\ee
where each factor
\be\label{Mobius}
{\mathfrak{M}}
(M_i)=\det\frac{(1+\tfrac{1}{\hbar}M_i)^{\tfrac12}(1-\tfrac{1}{\hbar}M_i)^{-\tfrac12} }
{\left( I_N \otimes I_N - \tfrac{1}{\hbar^2}
M_i\otimes M_i \right)^{\tfrac12}}
\ee
is responsible for the insertion of a Moebius strip, and each factor
\be\label{handle}
{\mathfrak{H}}(M_{i-1},M_{i})=
\det\left( I_N \otimes I_N - \tfrac{1}{\hbar^2}M_{i-1}\otimes M_{i} \right)^{-1}
\ee
is responsible for the insertion of a handle.
As for the geometric meaning of this factor as a function on a sphere with two holes, also see Appendix \ref{Mobius'}.

\br
We make a few comments:

(i) We can interpret the factor (\ref{handle}) as a sphere with two holes at the boundary of which the matrices 
$ M_i $ and $ M_ {i + 1} $ live. And the factor (\ref{Mobius}) can be interpreted as
an orientable covering of a projective sphere with a hole on whose boundary the matrix $ M_i $ lives (that is, as
sphere with two holes and an involution). Note that the formula (\ref{handle}) was already used in \cite{KMMM}
when describing the matrix model consisting of a (Hermitian) matrices $ M_1, \dots, M_n $  chain 
without any geometric interpretation.
(More precisely, in their model, the integrals of an expression  
$ {\mathfrak H}(M_1, M_2) {\mathfrak H}(M_2, M_3) \cdots $ were considered.)

 (ii) Note that in the right-hand sides of the equalities (\ref{MM1-all}), (\ref{MM1-all-klein}) and
(\ref{MM1-all-klein-handle}) contains the same factor $ {C} ({\Delta}^1, \dots, {\Delta}^{\ V}) $,
which depends only on the `` spectrum of stars ''. Moreover, the Hurwitz numbers in the right-hand sides
same set of branching profiles corresponding to these stars, and the difference is only in the covered
surfaces, these are respectively $ \Sigma $, $ \tilde \Sigma $ and $ \tilde {\tilde\Sigma} $.

(iii) The answer does not depend on how the matrices from the set $ \{M \} $ are distributed inside 
the integral - they can be rearranged: the left hand side of
(\ref{MM1-all-klein-handle}) depends
only how many matrices are spent on Moebius sheets and how much on handles.

(iv) What happens if we replace the three factors in the integral (\ref{MM1-all-klein-handle}) with tau functions, 
will be analyzed in another paper.

\er

\br\label{isospectral}
Note that any isospectral deformations of the set of $ \V $ monodromies (\ref{star-monodromy}) 
do not change the values of the intergals that we consider.
\er

\br\label{hypergeom}
The case when all the monodromies of stars $ W^*_i, \, i = 1, \dots, \V-1 $ are degenerate matrices is 
interesting in that in this case we are dealing with an integral over rectangular
matrices. If $ \Sigma = \mathbb{S}^2 $ and in addition
the monodromies of all stars except one or two (let it be $ W^*_i, \, i = 1,2 $) have spectra
$ {\rm Spect} W_i^* = (1,1, \dots, 1,0,0 \dots) $, $ {\rm rank} W_i^* = n_i $,
then for any $ \fa $ ($ \fa-n + \V = 2 $) the integral (\ref {MM1-all}) is
$$
\sum_{d\ge 0} (\hbar N)^{-nd}
 \sum_ {\lambda\atop |\lambda|=d} s_\lambda(W_1^*)s_\lambda(W_2^*)\prod_{i=3}^\V (n_i)_\lambda ,
$$
where $(x)_\lambda=(x)_{\lambda_1}(x-1)_{\lambda_2}\cdots (x-\ell+1)_{\lambda_\ell}$ is the Pochhammer symbol.
This sum is an example of the KP (see \cite{Sato},\cite{JM}) and TL (see \cite{Takasaki-Schur})  hypergeometric 
tau function \cite {OS-2000}. 
The spectrum of stars $ W_{1,2}^* $
in this case is called a set of Miwa variables.

And the integral (\ref {MM1-all-klein}) and with one insertion of a factor
(\ref{Mobius}), that is, if $ {\tilde \Sigma} = \mathbb{RP}^2 $ is equal to
$$
\sum_{d\ge 0} (\hbar N)^{-nd}
 \sum_{\lambda\atop |\lambda|=d}s_\lambda(W_1^*)\prod_{i=2}^\V (n_i)_\lambda
$$
This is also a hypergeometric tau function \cite{OST-I} for the hierarchy introduced in \cite{KvdLbispec}
which is not the KP one. Both tau functions are treated as formal series in $\hbar$.
\er

\subsection{Combinatorial meaning of the matrix model\label{combinatorics} }

To a graph whose all faces are homeomorphic to a disk, 
we can associate the permutation group $S_{2n}$, where $n$ is the number of edges and, respectively, $2n$ is the 
number of half-edges. To do this, all edges should be numbered on both sides with numbers from $1$ to $n$. 
The unordered set of these numbers is denoted by $ J $.
After that, cycles corresponding to faces are selected in the permutation group: the set of edge numbers read in 
the positive direction while walking around the capital of the selected face. 
We number faces. Let the face with the number $ m $  ($m=1,\dots,,\fa$) be associated with the cycle 
$f_m$.

For the first face with the total number of sides $ k_1 $, let the side labels be the numbers
$ i_1, i_2, \dots, i_k \in J $. The corresponding cycle in the group $ S_{2n} $ we denote
\be
(i_1,\dots, i_{k_1})
\ee
The monodromy of the face is precisely built on this cycle:
\be
Z_{i_1}C_{i_1}\cdots Z_{i_{k_1}}C_{i_{k_1}}
\ee
It is natural to call such matrix products constructed over a cycle as cycle products, but we
we will call them ``dressing  the cycle with matrices''.

The set of all cycles of $S_{2n}$ that correspond to the faces will be denoted as follows:
\bea\label{face-cycles}
f_1=(i_1,\dots, i_{k_1}),
\\
f_2=(i_{k_1+1},\dots,i_{k_1+k_2}),
\\
\dots\dots\dots 
\\
f_\fa=(i_{k_1+\cdots +1},\dots,i_{k_1+\cdots +k_\fa}),
\eea
where  the set of different numbers $i_1,\dots,i_{k_1+\cdots +k_\fa}=i_{2n}$ forms the set $J$, $|J|=2n$.

Traces of the monodromies of the faces that correspond to these cycles are
\bea\label{face-cycles-monodromies}
 \tr (M_1)=\tr\left(Z_{i_1}C_{i_1}\cdots Z_{i_{k_1}}C_{i_{k_1}} \right) 
 =:{\cal D}_Z\left[ f_1  \right],
\\
 \tr ( M_2) =\tr\left(Z_{i_{k_1+1}}C_{i_{k_1+1}}\cdots Z_{i_{k_1+k_2}}C_{i_{k_1+k_2}} \right)
 =:{\cal D}_Z\left[ f_2  \right],
\\
\dots \dots \dots
\\
\tr (M_\fa) = \tr\left(Z_{i_{k_1+\cdots +1}}C_{i_{k_1+\cdots +1}}\cdots 
Z_{i_{2n}}C_{i_{2n}} \right)=:{\cal D}_Z\left[ f_\fa  \right]
\eea
respectively. The operation ${\cal D}_Z[{\rm cycle}]$ we call the dressing of the cycles which denotes the replacement 
of the cycle by the trace of the product of the matrices $Z_xC_x$ whose 
numbers $x$ are equal to the numbers in the cycle.

We also introduce cycles related to vertices: each vertex is associated with the set of numbers  of the sides that
occur when approaching the edge, when we go around the vertex in the negative direction (clockwise).
We number the vertices. The vertex (star) with number $ s $ corresponds to the cycle 
$ \sigma_s $, $ s = 1, \dots, \V $ 
and the trace of the monodromy, related to this vertex:
\bea\label{star-cycles}
\sigma_1=(j_1,\dots, j_{s_1})
\\
\sigma_2=(j_{s_1+1},\dots,j_{s_1+s_2})
\\
\dots\dots\dots 
\\
\sigma_\V=(j_{s_1+\cdots +1},\dots,j_{2n})
\eea
where  $j_1,\dots,j_{2n}$ are different and belong to the set $J$, $|J|=2n$.

Traces of the monodromies of the vertices (stars) that correspond to these cycles are
\bea
 \tr (W^*_1)=\tr\left(C_{j_1}\cdots C_{j_{s_1}} \right)  =: {\cal D}\left[ \sigma_1 \right]
\\
 \tr ( W^*_2) =\tr\left( C_{j_{s_1+1}}\cdots C_{j_{s_1+s_2}} \right) =: {\cal D}\left[ \sigma_2 \right]
\\
\dots \dots \dots
\\
\tr (W^*_\V) = \tr\left(C_{j_{s_1+\cdots +1}}\cdots 
C_{j_{2n}} \right) =: {\cal D}\left[ \sigma_\V \right]
\eea
respectively. The operation $D[{\rm cycle}]$ we call the dressing of the cycles which denotes the replacement 
of the cycle by the trace of the product of the matrices $C_x$ whose 
numbers $x$ are equal to the numbers in the cycle.

By agreement, the edge with the number $ i $ ($ i = 1, \dots, n $) is matched with two numbers $ i $ and $ -i $, w
hich are assigned to different sides  of the ribbon edge. Assign to each edge $ i $ the transposition 
$ \alpha_i $, which permutes these two numbers.

For any graph $ \Gamma $ there is a remarkable relation \cite{ZL}
\be\label{combinatorial-graph}
\prod_{i=1}^n \alpha_i \prod_{i=1}^\fa f_i = \prod_{i=1}^\V\sigma_i
\ee
We can say that the involution $\prod_{i=1}^n \alpha_i $  takes the cycles of the faces of a graph into the cycles of 
the faces of the dual graph.

Graphs drawn on the covered surface in accordance with equation (\ref{combinatorial-graph}) are called children's 
drawings.

\br\label{half-edge}
 We return to the beginning of the chapter \ref{integral}, and recall that on both sides of the edge $ i $ ($ i> 0 $)
 in the graph $ \Gamma $ we placed the indices $ i $ and $ -i $ on the both sides of the ribs and that the 
 oriented sides of the rib ribbons we depicted 
 with arrows, i.e. the numbers $ i $ and $ -i $ were assigned to the arrows. Let's place
indexes are not just on the side of the edge where the arrow is, but also at the beginning of the arrow. 
Then the numbers $ i $ and $ -i $ number the so-called half-edges and the transposition 
$ \alpha_i, \, i> 0 $ is responsible for the transposition  of the half-edges.
\er

We turn to the integral (\ref{d=1}) and write in form
\be
\int {\cal D}_Z\left[ \prod_{i=1}^\fa f_i \right] d\Omega(Z) = {\cal D}\left[  \prod_{i=1}^\V\sigma_i \right]
\ee
The integrand is the sum of a large number of monomials consisting of the 
products of the entries of the matrices $\{Z\}$ and  $\{C\}$. Thanks to Gaussian integration, only members 
containing $(Z_i)_{a,b}(Z_{-i})_{b,a}(C_i)_{b,x}(C_{-i})_{a,y}$ are of importance, 
moreover, as a result of integration from this expression, only the product $\frac 1N (C_i)_{b,x}(C_{-i})_{a,y}$ 
remains:
\be\label{erasing-ribbon}
(Z_i)_{a,b}(Z_{-i})_{b,a}(C_i)_{b,x}(C_{-i})_{a,y} \quad  \to \quad \frac 1N (C_i)_{b,x}(C_{-i})_{a,y}
\ee
Recall that the matrix $Z_i$ ($i=1,\dots,n$) is depicted by an arrow, which is the side of the ribbon $|i|$, 
and the other side of this ribbon corresponds to the matrix $Z_{-i}$, which is depicted by an oppositely 
directed arrow; and the directions of the arrows are in the accordance with the positive orientation of the faces.

On the graph $\tig$, each solid arrow continues with a dashed arrow with the same number. Therefore, as follows from 
(\ref{erasing-ribbon}), after taking the integral, the dotted arrows around the small disk correspond to the product
of the source matrices along the sectors of the boundary of this disk (we get the monodromy matrix), the 
closure of the arrows corresponds to the trace of the monodromy of the star.
As a result of integration over all matrices $\{ Z \}$, it leads to the construction of monodromies of all stars, 
which proves formula (\ref{d=1}). The factor $N^{nd}$ in (\ref{d=1})  comes from the factor 
$1/N$ in the formula (\ref{erasing-ribbon}). You can visualize it this way: we erase all the ribbons and only 
small disks remain (stars).

We associate cycles with the traces of the product of matrices.

Consider the integrand in the formula (\ref {MM1-all'}). We want to number all matrix products
$ Z_iC_i $ for some fixed $ i $, and we will do it this way: we will number  these matrices from left to right and put
this number as superscripts: if the matrix met for the first time, we write $ Z_i^{(1)} C_i^{(1)} $ and so on
passing from left to right along the entire integrand (\ref{MM1-all'}).
We will do the same with the right-hand side of the equality (\ref{MM1-all'}); in this case we fix $ i $ and number 
the matrices $ C_i $: like $ C_i^{(1)}, C_i^{(2)}, \dots $
when we go from left to right along the entire right-hand side (\ref{MM1-all'}).
 There is arbitrariness in this procedure, because we have the freedom to rearrange matrices under the trail sign and 
 the freedom to rearrange the matrix traces themselves; 
we fix this arbitrariness choosing this order "by hands" .

Further, the integrand on the left side of the equality (\ref{MM1-all'}) can be associated with a set of 
cycles in the group $ S_{2nd} $. The right-hand side of the equality (\ref{MM1-all'}) can also be associated 
with a set of cycles in the group $ S_{2nd} $.
The reverse procedure can be called the dressing the cycle and will be denoted by the symbol 
$ {\cal D}_Z [{\rm cycle}] $ for dressing cycles on the left side (\ref{MM1-all'})
and the $ D [{\rm cycle}] $ symbol for dressing
cycles on the right side (\ref{MM1-all'}). On the left side, we replace each index $ i^{(a)} $ by the matrix
$ Z_i^{(a)}C_i^{(a)} $. On the right side, we replace each index $ j^{(a)} $ with the matrix $ C_j^{(a)} $.
By procedure of the dressing of a set of cycles, we mean the product of the  dressed cycles: 
${\cal D}\left[ f g \right]= {\cal D}\left[ f \right] {\cal D}\left[ g \right]$ and 
${\cal D}_Z\left[ f g \right]= {\cal D}_Z\left[ f \right] {\cal D}\left[ g \right]$.

Denote
\bea\label{face-cycles-cover}
f_1^{(1^d)}=  \left(i_1^{(1)},\dots,i_{k_1}^{(1)}   \right) \cdots 
\left(i_1^{(d)},\dots,i_{k_1}^{(d)}   \right)
\\
f_2^{(1^d)}=   \left(i_{k_1+1}^{(1)},\dots,i_{k_1+k_2}^{(1)}   \right)
 \cdots 
 \left(i_{k_1+1}^{(1)},\dots,i_{k_1+k_2}^{(1)}   \right)
\\
\dots\dots\dots 
\\
f_\fa^{(1^d)}=   \left(i_{k_1+\cdots +1}^{(1)},\dots,i_{2n}^{(1)}\right)
 \cdots 
\left(i_{k_1+\cdots +1}^{(d)},\dots,i_{2n}^{(d)}\right)
\eea
where $(1^d)$ denotes the partition $(1,1,\dots,1)$ consisting of only units.
Why it is convenient for us to label cycles with Young diagrams will become clear later.
The right-hand sides of these equalities include all possible 
$ i_j^{(a)} \, j = 1, \dots, 2n, \, a = 1, \dots, d $ with a total number of $ 2nd $.

Then we have
\be\label{face-cycles-cover-monodromy}
\left(\tr M_i \right)^d= {\cal D}_Z\left[ f_i^{(1^d)}
\right],\quad i=1,\dots,\fa
\ee
(compare to (\ref{face-cycles-monodromies})).
The cycles on the right-hand side can be treated as an unramified $ d $-sheeted covering of cycles in
on the right side in (\ref{face-cycles}): each of the indices $ i_j^{(a)} $ ($ a = 1,  \dots, d $) with its 
own neighborhood inside each of the cycles (cycles are labeled with $ a $) is projected onto $ i_j $.

Cycles can also be matched with polygons with numbered sides.
Then the polygons on the base surface $\Sigma$ cut along the edges of the graph $\Gamma$ 
are compared to the cycles (\ref{face-cycles}).
And the cycles (\ref{face-cycles-monodromies}) are matched by polygons
covering them in the amount of $ d $ copies
(picture: barbecue from strung polygons, skewers set vertically).

The integral (\ref{MM1-all'}) glues the polygons from the 'barbecue' to get the surface that covers $ \Sigma $,
exactly as the integral (\ref{d=1}) glues together
base surface of polygons (\ref{face-cycles}).

However, now we have a choice of which polygons to stick together, since each
the matrix $ Z_i $ from the set $ \{Z \} $ occurs $ d $ times: they are marked
as $ Z_i^{(1)}, \dots, Z_i^{(d)} $.
Wick's theorem is responsible for pairing.
And as a result, we get the sum on the right side of the equality (\ref{MM1-all'}).

Now, for each pairing, the transposition is responsible, which indicates which matrices are glued to which.
We denote these transpositions by $ \alpha_i^{(a, b)} $, where the superscript indicates that the matrix 
$ Z_i^{(a)} $ is paired
with the matrix $ Z_{- i}^{(b)} $.
Gaussian integral (\ref{MM1-all'}) is responsible for the complete pairing of all $ 2nd $ matrices
$ Z_i^{(a)}, \, i = 1, \dots, d \, i = 1 \dots, 2n $. This leads to the replacement of the products 
$ \prod_{i = 1}^n \alpha_i $ with such a sum
\be\label{Wick-transpositions}
 \sum_{w_1,\dots,w_\fa\in S_d}\hat{\alpha}(w_1,\dots,w_n),
\ee 
 where
\be\label{glue} 
 \hat{\alpha}(w_1,\dots,w_n):=     \prod_{a=1}^d\prod_{i=1}^n \alpha_i^{(a,w_i(a))},
\ee
where the permutations $ w_i $ is responsible for the Wick rule: these permutations 
correspond to all possible pairings. We call the element (\ref{glue}) of the group $ S_{2nd} $ 
the gluing element. Formula (\ref{Wick-transpositions}) describes the summation over all gluings. 
The involution $ \prod_{i = 1}^n \alpha_i $ of the formula (\ref{combinatorial-graph}) corresponding 
to the edges of the graph $ \Gamma $
is a special case of (\ref{Wick-transpositions}) obtained for $ d = 1 $.

Now recall a fact about the composition of a transposition with a product of disjoint cycles.
A transposition permutes some two elements. There are two cases. In the first, both elements belong to one cycle.
Then the product of the transposition and the cycle will give two disjoint cycles. In the second case - two 
rearranged elements lie in two different disjoint cycles, the action of transposition combines them into one.
This property is known as "cut-or-join action".
If we multiply the gluing operator (\ref{glue}) by the product of all cycles from the set of the ``barbecue cycles'' 
on the right side (\ref{face-cycles-monodromies})
it turns out a new set of cycles.
Note that the resulting set of cycles can be interpreted as the cycles,
covering the original set (\ref{star-cycles}), since each element of $ \alpha_i^{(a, b)} $ acts
over its edge $ i $ of the graph $ \Gamma $.

Consider a vertex, for example, the vertex number $ 1 $, from which $ s_1 $ edges come out, 
see (\ref{star-cycles}). Let covering cycles have lengths $ s_1 \Delta^1_1 \ge \cdots \ge s_1 \Delta^1_\ell $
(the sum of all lengths should be equal $ 2nd $. The cycles will be denoted respectively
$ \sigma_1^{(\Delta^1_1)}, \dots, \sigma_1^{(\Delta^1_\ell)} $, where the cycle 
$ \sigma_1^{(\Delta^1_p)} $ is the $ \Delta^1_p $ -listed
cover of the cycle (\ref{star-cycles}). The first two look like this:
  \bea\label{star-cycles-cover}
\sigma_1^{(\Delta^1_1)}=\left(j_1^{(1)},\dots, j_{s_1}^{(1)},j_1^{(2)},\dots, j_{s_1}^{(2)},\dots 
j_1^{(\Delta^1_1)},\dots, j_{s_1}^{(\Delta^1_1)}
\right)
\\
\sigma_1^{(\Delta^1_2)}=\left(j_1^{(\Delta^1_1+1)},\dots, j_{s_1}^{(\Delta^1_1+1)},
j_1^{(\Delta^1_1+2)},\dots, j_{s_1}^{(\Delta^1_1+2)},\dots 
j_1^{(\Delta^1_1+\Delta^1_2)},\dots, j_{s_1}^{(\Delta^1_1+\Delta^1_2)}
\right)
\eea
 etc.
Denote
 $$
\sigma_s^{\Delta^s} := \prod_{m=1}^{\ell(\Delta^i)}\sigma_i^{\Delta^s_m}.
$$
The set $\{\Delta^s=(\Delta^s_1,\Delta^s_2,\dots), \,s=1,\dots,\V\}$ 
is a set of Young diagrams that we we will attribute the corresponding vertices of the graph 
$ \Gamma $ (to the stars of the graph $ \tig $).

Each trace of the degree of the matrix corresponds to a covering cycle.
We have
\be\label{covering-cycle}
\sigma_i^{\Delta^i_1}\cdots \sigma_i^{\Delta^i_{k_1}}
\,\,\,\longleftrightarrow \,\, \,
\tr (W_i^*)^{\Delta^i_1}\cdots \tr (W^*_i)^{\Delta^i_{k_1}}\,
=: \,{\cal D}[\sigma_i^{\Delta^i_1}\cdots \sigma_i^{\Delta^i_{k_1}}]
\ee
 
 \br
 For the Feynman graph $ \tig $, instead of a vertex, we should consider the boundary of the small disk,
 consider it a polygon, with sides consisting of $ s_1 $ dashed segments border (segment connects
 adjacent outgoing edges of the ribbon), and thus considering the covering this polygon with a system of 
 polygons with the number of sides $ s_1 \Delta^1_1, \cdots, s_1 \Delta^1_\ell $.
\er

Choose a set of permutations $ w_1, \dots, w_n $ and consider the equation
\be\label{Hurwits-number-equation}
\hat{\alpha}(w_1,\dots,w_n)\prod_{i=1}^\fa f_i^{(1^d)}
= \prod_{i=1}^\V\sigma_i^{\Delta^i}
\ee
with indeterminants $ \Delta^1, \dots, \Delta^\V $,
in which the right-hand side describes what is the result of the composition of the involution 
(\ref{glue}) with the product of the
cycles  $ f_1^{(1^d)}, \dots, f_\fa^{(1^d)} $ (which are treated as preimiges of the  cycles $ f_1, \dots, f_\fa $).
It's easy to understand that the set of cycles
$\{\sigma_i^{\Delta^i_m},\,i=1,\dots,n,\,m=1,\dots,\ell(\Delta^i)\}$
is the set of disjoint cycles, this follows from the properties of $ \alpha_i^{(a, b)} $.

Equation (\ref{Hurwits-number-equation}) corresponds to a Wick pairing given by a given set $w_1,\dots,w_n$
and corresponds to a given Feynman graph.
It  plays the role of the equation 
(\ref{combinatorial-graph}), but for 
another graph: the graph $ \hat\Gamma $, which covers the graph $ \Gamma $. 
Graph $ \hat\Gamma $ with punctured vertices is the $ d $ -listed and unramified cover of the graph 
$ \Gamma $ with punctured vertices. The vertices of $ \Gamma $ are branch points with profiles 
$ \Delta^1, \dots, \Delta^\V $.

Now consider a given set of Young diagrams $ \Delta^1, \dots, \Delta^\V $ of the same weight $ d $ , and assume that
(\ref{Hurwits-number-equation}) is the equation for the unknown $ w_1, \dots, w_n $.
Divide the number of solutions of this equation by the number $ (d!)^N $.
 This is the Hurwitz number $ H_\Sigma(\Delta^1, \dots, \Delta^\V) $. It describes how many ways to glue a covering of
  the surface $ \Sigma $ from polygons  if all branch profiles are specified.
The division by $ (d!)^N $ is natural, since in our construction the covering of each cycle 
$ f_i, \, i = 1, \dots, n $
consists of $ d $ identical cycles (polygons). This corresponds to the geometric definition.
 Hurwitz numbers as the number of nonequivalent coverings for a given set of branching profiles, 
 see Appendix \ref{Definition}.

The matrix integral (\ref{MM1-all'}) we write as follows: 
\be\label{MM1-all''}
 \left(\frac{N^d}{d!}\right)^{n} \int {\cal D}_Z\left[ \prod_{i=1}^\fa f_i^{(1,1,\dots,1)} \right]d\Omega =
 \sum_{\Delta^1,\dots,\Delta^\V} H_\Sigma(\Delta^1,\dots,\Delta^\V)
{\cal D}\left[ \prod_{i=1}^\V\sigma_i^{\Delta^i} \right]
\ee
which is a manifistation of the  equality
\be\label{Hurwitz-combinatorial}
\sum_{w_1,\dots,w_\fa\in S_d} \hat{\alpha}(w_1,\dots,w_n)\prod_{i=1}^\fa \frac{f_i^{(1,1,\dots,1)}}{d!}
= \sum_{\Delta^1,\dots,\Delta^\V} H_\Sigma(\Delta^1,\dots,\Delta^\V) \prod_{i=1}^\V\sigma_i^{\Delta^i}
\ee

This equality describes how the higher Feynman graphs of the integral (\ref{MM1-all}) cover the lower graph 
$ \ tig $. The types of the preimages of the small disks of the graph $ \tig $) are given by the 
set $ \Delta^1, \dots, \Delta^\V $.

If instead of the cycles $f^{(1,1,\dots,1)}_i$ we take $ f^{\tilde{\Delta}^i}_i $, then instead of
(\ref{Hurwitz-combinatorial}) and of
(\ref{MM1-all'}) we get respectively
\be\label{Hurwitz-combinatorial-tilde}
 \sum_{w_1,\dots,w_\fa\in S_d} \hat{\alpha}(w_1,\dots,w_n)
 \prod_{i=1}^\fa \frac{f_i^{\tilde{\Delta}^i}}{z_{\tilde{\Delta}^i}}
\ee 
\be 
= 
\sum_{\Delta^1,\dots,\Delta^\V} 
H_\Sigma(\tilde{\Delta}^1,\dots,\tilde{\Delta}^\fa,\Delta^1,\dots,\Delta^\V) \prod_{i=1}^\V\sigma_i^{\Delta^i}
\ee
where the number $z_\Delta$  is defines in Appendix by (\ref{z_Delta}),
and
\be\label{MM1-all'''}
N^{nd} \int {\cal D}_Z\left[ \prod_{i=1}^\fa \frac{f_i^{\tilde{\Delta}^i}}{z_{\tilde{\Delta}^i}}
 \right]d\Omega
\ee
\be 
 =\sum_{\Delta^1,\dots,\Delta^\V}
 H_\Sigma(\tilde{\Delta}^1,\dots,\tilde{\Delta}^\fa,\Delta^1,\dots,\Delta^\V)
{\cal D}\left[ \prod_{i=1}^\V\sigma_i^{\Delta^i} \right]
\ee
The Hurwitz numbers on the right-hand side contain two sets of profiles: one set corresponds to
a set of preimages of faces of the graph $ \Gamma $ (a set of preimages of the main loops of the 
Feynman graph $ \tig $),
and another set corresponds to the set of  preimages of the vertices of the graph $ \Gamma $ 
(or, what is the same, to the set
the preimages of the small disks of the graph $ \tig $) when a high Feynman graph covers the lowest Feynman
graph $\tig$. The position of the branch points does not affect
on the Hurwitz numbers, but for clarity, we can assume that the branch points are located in the capitals of the
 faces and the vertices of the graph $ \Gamma $. Or, what's the same thing - at the vertices of the graph dual to 
Feynman's graph $ \tig $.

Let us write down the combinatorial relation of a Feynman diagramm related to
(\ref{MM1-all-klein-handle}). 
Instead of (\ref{Hurwits-number-equation}), we obtain
\be
\sum_{w_1,\dots,w_\fa\in S_d} \hat{\alpha}(w_1,\dots,w_n)\prod_{i=1}^{\fa_1} \frac{f_i^{(1,1,\dots,1)}}{d!}
\prod_{i=\fa_1+1}^\fa \left(\sum_{\mu_i \in {\cal P} } f_i^{(\mu_i)} D(\mu_i)\right) 
\ee
$$
=
\sum_{\Delta^1,\dots,\Delta^\V} H_{\tilde \Sigma}(\Delta^1,\dots,\Delta^\V) \prod_{i=1}^\V\sigma_i^{\Delta^i}
$$
where ${\cal P}$ is the set of all partitions and where  $D(\mu)$ is written down in Appendix 
\ref{Hurwitz numbers and representation theory}.

\be
\label{MM1-all-klein-handle'} 
 \sum_{w_1,\dots,w_\fa\in S_d} \hat{\alpha}(w_1,\dots,w_n)\prod_{i=1}^{\fa_1} \frac{f_i^{(1,1,\dots,1)}}{d!}
\prod_{i=\fa_1+1}^{\fa_2}{\mathfrak{m}}(f_i)
\left(\prod_{i=\fa_2+1}^{\fa-k}
{\mathfrak{h}}(f_i,f_{i+1})\right)
\ee
$$
=
\sum_{{\Delta}^1,\dots , {\Delta}^{\V}\atop |\Delta^1|=\cdots = |\Delta^{\V}|=d} 
H_{\tilde{\tilde\Sigma}}({\Delta}^1,\dots , {\Delta}^{\V})
{C}({\Delta}^1,\dots , {\Delta}^{\V} ),
$$
where ${\cal P}$ is the set of all partitions, and  $D(\mu)$ and $\frac{1}{z_{\mu}}$ are given in Appendix 
\ref{Hurwitz numbers and representation theory} and 
where the factors
\be\label{Mobius''}
{\mathfrak{m}}(f_i)= \left(\sum_{\mu_i \in {\cal P} } f_i^{(\mu_i)} D(\mu_i)\right) 
\ee
correspond to the insertition of M\"obius strips, and factors
\be\label{handle'}
{\mathfrak{h}}(f_i,f_{i+1})=
\left(\sum_{\mu_i \in {\cal P} } f_i^{(\mu_i)}f_{i+1}^{(\mu_i)} \frac{1}{z_{\mu^i}}\right) 
\ee
correspond to the insertion of handles.

\section{Discussion \label{Discussion}}

\subsection{Differential operators}

(i)
One can interpret the Gaussian integral as the integral of $ n $ -component two-dimensional charged bosonic fields
$Z_i$ and $Z^\dag_i$:
$$
\int (Z^\dag_i)_{a,b}(Z_j)_{b',a'}
d\Omega =
<(Z^\dag_i)_{a,b}(Z_j)_{b',a'}>=\frac 1N \delta_{a,a'}\delta_{b,b'}\delta_{i,j},
$$
for $i,j=1,\dots,n,\quad a,b=1,\dots, N$.

The Fock space of these fields is all possible polynomials from the matrix elements of the matrices
$Z_1,\dots,Z_n$.

The operators $ (Z_i)_{ab} $ can be considered creation operators, and the operators
\be\label{correspondence}
({Z}^\dag_i)\,\to\,\frac 1N \partial_i,\quad   (\partial_i)_{a,b}=\frac 1N \frac{\partial}{\partial Z_{b,a}}
\ee
- ellimination operators that act in this space.
The integrands in (\ref{MM1}) should be considered anti-ordered, that is, all ellimination operators (all derivatives)
considered to be moved to the left, while the matrix structure is considered to be preserved. We will denote
this is anti-ordering of some $ A $ by the symbol $ :: A :: $, where $ A $ is a polynomial of matrix elements
of the matrices

From this point of view, on different sides of the ribbon of $\tig$ with the number $i$ we place the canonically 
conjugated coordinates $Z_i$ and momenta $\partial_{Z_i}$.

We recall that all partions throught the paper have the same weight $d$.

(ii)
Then, for example, the relation (\ref{MM1-tau}) takes the form
\begin{equation}\label{Operator-tau}
:: \tau_g(M_1,\dots,M_{\fa})::
=\sum_{d=0}^\infty
\sum_{{\Delta}^1,\dots , {\Delta}^{\V}\atop |\Delta^1|=\cdots = |\Delta^{\V}|=d} 
H_{\Sigma'}(g|{\Delta}^1,\dots , {\Delta}^{\V})
{C}({\Delta}^1,\dots , {\Delta}^{\V} ),
\end{equation}
which is written more compact than (\ref{MM1-tau}).

We give another relation:
\be
\left(::\prod_{i=1}^\fa s_{\lambda^i}(M_i)::\right)\cdot 1 = 
\delta_{\lambda^1,\dots,\lambda^\fa} \left(\frac{{\rm dim}\,\lambda}{d!}  \right)^{-n} \prod_{i=1}^\V s_\lambda(W_i^*)
\ee
where $\lambda^1,\dots,\lambda^\fa=\lambda$  is a set of Young diagrams, and where $\delta_{\lambda^1,\dots,\lambda^\fa}$ 
is equal 1, if $\lambda^1=\dots =\lambda^\fa=\lambda$ and is equal to 0 otherwise. 

It looks like a simple rewrite, but can be helpfully used. Let us derive a beautiful formula 
(Theorem 5.1 in \cite{MM3}), namely (\ref{MM3}) (and see also 
articles \cite{Olshanski-19},\cite{Olshanski-199},\cite{Okounkov-1},\cite{Okounkov-19},\cite{Okounkov-199},
\cite{Okounkov-1996}).

In order to do this we should use the freedom to choose the source matrices:

(iii)
For a partition $\Delta=(\Delta_1,\dots,\Delta_\ell)$ and a face monodromy $M_i$ and a star monodromy $W^*_i$,
let us introduce notations
\be
{\cal M}_i^{\Delta^i}=\tr \left((M_i)^{\Delta^i_1}\right)\cdots \tr \left((M_i)^{\Delta^i_\ell}\right)
\ee
\be
{\cal C}_i^{\Delta^i}=\tr \left((W_i^*)^{\Delta^i_1}\right)\cdots \tr \left((W^*_i)^{\Delta^i_\ell}\right)
\ee

Then (\ref{MM1-all'''}) is written as
\be\label{MM1-all'''matrices}
N^{nd} \int \left[ \prod_{i=1}^\fa \frac{{\cal M}_i^{\tilde{\Delta}^i}}{z_{\tilde{\Delta}^i}}
 \right]d\Omega
\ee
\be 
 =\sum_{\Delta^1,\dots,\Delta^\V}
 H_\Sigma(\tilde{\Delta}^1,\dots,\tilde{\Delta}^\fa,\Delta^1,\dots,\Delta^\V)
\left[ \prod_{i=1}^\V {\cal C}_i^{\Delta^i} \right]
\ee

Let us write the most general generating function for Hurwitz numbers which was obtained in \cite{NO2020}:
\be
\label{full-generating function}
N^{nd}\int \left[ \prod_{i=1}^{\fa_1} \frac{{\cal M}_i^{\tilde{\Delta}^i}}{z_{\tilde{\Delta}^i}}
 \right] 
 \left(
\prod_{i=\fa_1+1}^{\fa_2}{\mathfrak{M}}(M_i) \right)
\left(\prod_{i=\fa_2+2,\fa_2+4,\dots}^{\fa}
{\mathfrak{H}}(M_{i-1},M_{i})\right)
  d\Omega
\ee
\be \label{full-rhs}
=
\sum_{{\Delta}^1,\dots , {\Delta}^{\V}\atop |\Delta^1|=\cdots = |\Delta^{\V}|=d} 
H_{\tilde{\tilde\Sigma}}(\tilde{\Delta}^1,\dots , \tilde{\Delta}^{\fa_1},{\Delta}^1,\dots , {\Delta}^{\V})
{C}({\Delta}^1,\dots , {\Delta}^{\V} ),
\ee
where
\be\label{Hurwitz-general}
 H_{\tilde{\tilde\Sigma}}(\tilde{\Delta}^1,\dots , \tilde{\Delta}^{\fa_1},{\Delta}^1,\dots , {\Delta}^{\V})=
 \sum_{\lambda\in {\cal P}} 
 \left( \frac{{\rm dim}\,\mu}{d!} \right)^{\fa-n+V-2h-m}
 \varphi_\mu(\tilde{\Delta}^1)\cdots \varphi_\mu(\tilde{\Delta}^{\fa_1})
 \varphi_\mu({\Delta}^1)\cdots \varphi_\mu({\Delta}^{\V})
\ee
where $h=\frac12 (\fa -\fa_2)$ is the number of handles and $m=\fa_2-\fa_1$ is the number of
Moebius stripes glued to $\Sigma$ where the graph $\tig$  (modified dessins d'enfants) was drawn.
The Euler characteristic of ${\tilde{\tilde\Sigma}}$ is $\fa-n+V-2h-m=\fa_1-n+\V$. Hurwitz number
(\ref{Hurwitz-general}) counts the coverings of ${\tilde{\tilde\Sigma}}$ with branching profiles
$\tilde{\Delta}^1,\dots , \tilde{\Delta}^{\fa_1},{\Delta}^1,\dots , {\Delta}^{\V}$.

Let us multiply the both sides of (\ref{full-generating function}) by
$$
 \prod_{i=k+1}^{\fa_1}\frac{{\rm dim}\,\mu^i}{d!}\varphi_{\mu^i}(\tilde{\Delta}^i) z_{\tilde{\Delta}^i}
$$
(where $k\le\fa$), and then sum the both sides (\ref{full-generating function}) and (\ref{full-rhs}) over 
$\tilde{\Delta}^i,\,i=k+1,\dots,\fa_1$, taking into account 
\be
s_\mu(X)=\frac{{\rm dim}\,\mu}{d!}\sum_\Delta\varphi_{\mu}(\Delta){\cal X}^\Delta ,
\ee
when evaluating  (\ref{full-generating function}),
where
\be
{\cal X}^{\Delta}=\tr \left((X)^{\Delta_1}\right)\cdots \tr \left((X)^{\Delta_\ell}\right)
\ee
and the orthogonality relation (\ref{orth2}) when evaluating  (\ref{full-rhs}). We obtain
\be
\label{MM1-all-klein-handle''}
\int 
N^{nd} \left[ \prod_{i=1}^k \frac{{\cal M}_i^{\tilde{\Delta}^i}}{z_{\tilde{\Delta}^i}}
 \right] 
 \left(
\prod_{i=k+\mc+1}^{k+\mc+m}{\mathfrak{M}}(M_i) \right)
\left(\prod_{i=k+\mc+m+2,k+\mc+m+4,\dots}^{\fa}
{\mathfrak{H}}(M_{i-1},M_{i})\right)
\prod_{i=k+1}^{k+\mc} s_{\mu^i}(M_i)
  d\Omega
\ee
\be
=
\delta_{\mu^{k+1},\dots,\mu^{\fa-k} }
 \left( \frac{{\rm dim}\,\mu}{d!} \right)^{k-n-2h-m}
 \varphi_\mu(\tilde{\Delta}^1)\cdots \varphi_\mu(\tilde{\Delta}^k)
\left[ \prod_{i=1}^\V s_\lambda(W_i^*)) \right]
\ee
where $2h=\fa-\fa_1-m$ $\mc$

\br\label{chess-desk}
 Suppose that the edges of the graph $\Gamma$ can be painted like a chessboard in black and white faces so that 
 the face of one color borders only the faces of a different color. Then the matrices from the set $\{ Z^\dag\}$ 
 (i.e., differential operators) can be assigned to the sides of the edges of white faces, that is, the matrices 
 from the set $\{Z\}$ to the sides of black faces. In this case, the monodromies of the white faces will be those 
 differential operators which will act on the monodromy of black faces.
 
\er

The most natural and simple case is the following 'polarization': Suppose that $M_i,\,i=k+1,\dots,\fa_1$ 
are black faces and the rest part of the face monodromies $M_i$ are while faces (see Remark \ref{chess-desk}).

(I) Let $m=h=0$.
Take as a graph $\tig$ a child's drawing - sunflower with $n$ white petals drawn on the background of black night sky.
See (b) in the figure for $\Gamma$ with 2 petal as an example. 
There is 1 vertex of $\Gamma$
which inflated and we get a small disk as the center of sunflower. We have $n+1$ faces of $\Gamma$: $n$ petals, and the big 
and a big face, embracing all the petals and containing infinity.
Then we place all ``momentums'' inside the petals:
$$
M_i=C_{-i}Z_i^\dag,\,i=1,\dots,n .
$$. 
Then all ``coordinates'' (the collections of $\{(Z_i)_{a,b}\}$) are placed on 
the other side 
of the ribbons, they are places along the boundary of the big embracing black face: 
$$
M_{n+1}=Z_1C_1\cdots Z_nC_n
$$. 
Let remove the sign tilde above Young diagrams,
then, 
\be\label{petal-integral}
\int N^{nd} 
\left[ \prod_{i=1}^n 
\frac{{\cal M}_i^{\Delta^i}} {z_{\Delta^i}}
 \right]  
 s_{\mu}\left( Z_1C_1\cdots Z_nC_n \right)  d\Omega= 
 \varphi_\mu(\Delta^1)\cdots \varphi_\mu(\Delta^1) s_\mu(C_{-1}C_1\cdots C_{-n}C_n)
\ee
It is equivalent to
\be\label{vector-fields-action}
{\cal W}_{C_{-1}}^{\Delta^1}\cdots {\cal W}_{C_{-n}}^{\Delta^n}  \cdot s_\mu(Z_1C_1\cdots Z_nC_n)=
\varphi_\mu(\Delta^1)\cdots \varphi_\mu(\Delta^1) s_\mu(C_{-1}(Z_1)C_1\cdots C_{-n}(Z_n)C_n),
\ee
where each $ N \times N $  matrix $ C_{-i} $  can now depend, for example, polynomially on
 $Z_i$, $i=1,\dots, n$ and where
\be
{\cal W}_{C_{-i}}^{\Delta^i}=
:\tr\left( ( C_{-i}(Z_i)\partial_i  )^{\Delta^i_1} \right)\cdots 
\tr\left( ( C_{-i}(Z_i)\partial_i  )^{\Delta^i_\ell} \right):\,,
\ee
where each $C_{-i}\partial_i $ is a matrix whose entries are differential operators, more precisely, are
the following vector fields:
\be
(C_{-i}\partial_i)_{a,b}:=\sum_{c=1}^N (C_{-i})_{a,c}\frac{\partial}{\partial (Z_i)_{b,c}}
\ee
The normal ordering indicated by two dots is the same here as in \cite{MM3} - that is,
while maintaining the matrix structure, the derivative operators do not act on
$ C_{-i} = C_{-i}(Z_i) $. Note that the normal ordering procedure is necessary in order  
the equation (\ref {vector-fields-action}) was equivalent to the equality (\ref{petal-integral})!

The ordering is the same as in \cite{MM3}: keeping the matrix structure the derivatives do not 
act on $C_{-i}=C_{-i}(Z_i)$. 
Notice that the ordering is necessary to relate (\ref{vector-fields-action}) to (\ref{petal-integral})!

If we now take the case $ n = 1 $ (one petal) and, in addition, $ C_{- i} = Z_i $, then we get the 
desired formula (\ref {MM3}).

Take another example with the same graph $ \tig $
and with the same monodromies. However, let $ k = 0 $.
In this case the integral (\ref{petal-integral}) can be re-written as the relation
\be\label{handle-mob-schur}
\hat{\mathfrak M}_1\cdots \hat{\mathfrak M}_m
\hat{\mathfrak H}_1\cdots
 \hat{\mathfrak H}_h \cdot s_\mu(Z_1C_1\cdots Z_nC_n)=
\left(\frac{{\rm dim}\,\mu}{|\mu|!}\right)^{-2h-m-n} s_\mu(C_{-1}(Z_1)C_1\cdots C_{-n}(Z_n)C_n),
\ee
where
\be
\hat{\mathfrak M}_i=:
e^{\frac 12 \sum_{j>0} \frac 1j\left(\tr \left(C_{-i}\partial_i \right)^j\right)^2
+ \sum_{j>0,{\rm odd}}\frac{1}{j}
\tr \left(\left(C_{-i}\partial_i \right)^j\right)}: \,,\quad i=1,\dots,m ,
\ee
\be
 \hat{\mathfrak H}_i= :e^{\sum_{j>0}\frac 1j \tr\left(\left(C_{1-m-2i}\partial_{m+2i-1}\right)^{j}  \right) 
\tr\left(\left(C_{-m-2i}\partial_{m+2i}\right)^{j}  \right) }:\,,\quad i=1,\dots, h.
\ee
\br
When $ C_{- i} = Z_i, \, i = 1, \dots, n $ both equalities (\ref {vector-fields-action}) and 
(\ref {handle-mob-schur}) describe eigenvalue problems for the corresponding Hamiltonians in the two-dimensional 
bosonic theory.
Perhaps a comparison with the case analyzed by Dubrovin is appropriate. This is the case 
$ n = 1 $, $ \fa = 1 $, $ {\cal W}^{(n)} $.
In this case, the operators $ {\cal W}^{(n)} \,, n = 1,2, \dots $ are the dispersionless Hamiltonians
KdV equations \cite{Dubr}.
\er

\br
The case $C_{-i}$ does not depend is also interesting in case the monodromies of the stars are degenerate 
matrices, then the whole intergal is related to the integration over rectanguler matrices. As an example
one can choose $C_{-1}C_1 \cdots C_{-n}C_n$ in (\ref{petal-integral}) as $\diag (1,1,\dots,1,0,0,\dots,0)$.
Then we get the Pochhhamer symbol in the right hand side which allows to related the whole integral to the
hypergeometric tau function \cite{OS-2000}. It will be discussed in a more detailed text where we
plan to relate out topic to certain topics in 
 \cite{Olshanski-19},\cite{Olshanski-199},\cite{Okounkov-1},\cite{Okounkov-19},\cite{Okounkov-199},\cite{Okounkov-1996}.
\er

Another example. $\Gamma$ has 2 vertices which are connected by 4 edges. We 
$$
:\bpow_{\Delta^1}(\partial_1C_{-1}\partial_2C_{-2})
\bpow_{\Delta^2}(\partial_3C_{-3}\partial_4C_{-4}):  \left(
s_{\lambda}(Z_1C_1Z_4C_4)s_{\mu}(Z_2C_2Z_3C_3)\right)
$$
$$
= \delta_{\lambda,\mu} \left( \frac{{\rm dim}\,\mu}{d!} \right)^{-2}
\varphi_\mu(\Delta^1)\varphi_\mu(\Delta^2)
s_{\mu}(C_1C_{-4}C_3C_{-2})s_{\mu}(C_4C_{-1}C_2C_{-3})
$$
In particular, if one takes $C_3=C_4=C$ and 
$C_{-1}=C_{-1}(Z_2),\,C_{-2}=C_{-2}(Z_1),\,C_{-3}=C_{-3}(Z_3),\,C_{-4}=C_{-4}(Z_4) $ he gets 
$$
:\bpow_{\Delta^1}(C_{-2}(Z_1)\partial_1 C_{-1}(Z_2) \partial_2 )
\bpow_{\Delta^2}(\partial_3 C_{-3}(Z_{3})\partial_4 c_{-4}(Z_{4})):  
s_{\lambda}(Z_1 C_1 Z_{4}C )s_{\mu}( Z_{2} C_2  Z_3 C  )
$$
$$
= \delta_{\lambda,\mu}  \left( \frac{{\rm dim}\,\mu}{d!} \right)^{-2}
\varphi_\mu(\Delta^1)\varphi_\mu(\Delta^2)
s_{\mu}(C_{-2}(Z_1)C_1C_{-4}(Z_4)C)s_{\mu}(C_{-1}(Z_2)C_2C_{-3}(Z_3)C)
$$

In particular, if one takes $C_3=C_4=C$ and $C_{-1}=Z_2,\,C_{-2}=Z_1,\,C_{-3}=Z_3,\,C_{-4}=Z_4 $ (Euler fields) 
he gets an  eigenvalue problem:
$$
:\bpow_{\Delta^1}(Z_1\partial_1 Z_2 \partial_2 )
\bpow_{\Delta^2}(\partial_3 Z_{3}\partial_4 Z_{4}):  \left(
s_{\lambda}(Z_1C_1Z_4C)s_{\mu}(Z_2C_2Z_3C)\right)
$$
$$
= \delta_{\lambda,\mu}  \left( \frac{{\rm dim}\,\mu}{d!} \right)^{-2}
\varphi_\mu(\Delta^1)\varphi_\mu(\Delta^2)
s_{\mu}(Z_1 C_1 Z_{4}C )s_{\mu}( Z_{2} C_2  Z_3 C  )
$$
In case $C_{-1}=C_{-2}=C_{-3}=C_{-4}=\mathbb{I}_N$ we get 
$$
:\bpow_{\Delta^1}(\partial_1 \partial_2 )
\bpow_{\Delta^2}(\partial_3 \partial_4 ):  \left(
s_{\lambda}(Z_1 C_1 Z_{4}C_4 )s_{\mu}( Z_{2} C_2  Z_3 C_3  )\right)
$$
$$
= \delta_{\lambda,\mu}  \left( \frac{{\rm dim}\,\mu}{d!} \right)^{-2}
\varphi_\mu(\Delta^1)\varphi_\mu(\Delta^2)
s_{\mu}(C_1 C_3)s_{\mu}(C_2 C_4)
$$

Now we consider another example $n=4$ with 
the graph obtained from the graph (a) in the fig 2 and 3  drawn on the torus
by doubling the edges: instead of each edge we draw two ones. 
We have $\Gamma$ with one vertex, four edges and three faces
and obtain
$$
 :\bpow_{\Delta^1} \left(\partial_1 C_{-1}\partial_3 C_{-3}  \right)
 \bpow_{\Delta^1} \left(\partial_2 C_{-2}\partial_4 C_{-4}  \right):
 s_\lambda\left( Z_1 C_1  Z_2 C_2 Z_3 C_3   Z_4 C_4 \right)
$$ 
$$ 
 = \left( \frac{{\rm dim}\,\mu}{d!} \right)^{-2}
 s_\lambda\left(C_1 C_{-2} C_4 C_{-1} C_3 C_{-4} C_2 C_{-3}  \right)
$$
Take $C_{-1}=C_{-1}(Z_3),\,C_{-2}=C_{-2}(Z_2),\,C_{-3}=C_{-3}(Z_1),\,C_{-4}=C_{-4}(Z_4)$ and $C_2=C_4=C$.
As an example we obtain
$$
 :\bpow_{\Delta^1} \left(Z_1\partial_1 Z_3\partial_3   \right)
 \bpow_{\Delta^1} \left(\partial_2 Z_{2}\partial_4 Z_{4}  \right):
 s_\lambda\left( Z_1 C_1  Z_2 C Z_3 C_3   Z_4 C \right)
$$ 
$$ 
 =\left(\frac{{\rm dim}\,\lambda}{d!}   \right)^{-2}\varphi_\lambda(\Delta^1)\varphi_\lambda(\Delta^2)
 s_\lambda\left(Z_1 C_1 Z_{2} C Z_{3} C_3 Z_{4} C   \right)
$$

(iv)  Let us notice that if we take a dual graph to the sunflower graph with $n=1$ (dual to one petal $\Gamma$, which is just
a line segment, see fig 1 ), in this case we have one face and two vertices, we get a version of the Capelli-type relation.
Then it is a task to compare explicitly such relations with beautiful results \cite{Okounkov-1}, 
\cite{Okounkov-19}, \cite{Okounkov-199},\cite{Okounkov-1996}.

(v)  There are several allusions to the existence of interesting structures related to quantum integrability. 
First, as noted in \cite{NO2020}  by this appearance 2D Yang-Mills theory \cite{Witten}. See also
possible connection to \cite{Gerasimov-Shatashvili}.  Then the appearance of 
the Yangians in works \cite{Olshanski-19},\cite{Olshanski-199} which, we hope, can be related to our subject. 
And finally, the work \cite{Dubr}.

(vi) There is a direct similarity between integrals over complex matrices and integrals over unitary matrices.
However, from our point of view direct anologues of the relations in the present paper are more involved in
the case of unitary matrices. In particular, Hurwitz numbers are replaced by a special combination of these numbers.

\subsection{Comparison with the model of Hermitian matrices}

For comparison, we write the famous single-matrix model \cite {BrezinKazakov}, \cite{GrossMigdal} with 
the added source matrix:
$$
{\cal Z}_N(\bpow,C)=\int e^{N\sum_{m>0} \frac Nm p_m \tr \left( (XC)^m \right)}e^{-N\tr \left(X^2\right)}
\prod_{a=1}^N d X_{a,a}\prod_{a>b}^N d\Re(X_{a,b})d\Im(X_{a,b}),
$$
where $ C $ is the $ N \times N $ matrix, and $ X $ is the $ N \times N $ Hermitian matrix.
Such a model was considered in \cite{Kazakov}.

This model can also be considered as a generating function for coverings of the graph $ \Gamma $ on $ \mathbb{S}^2 $ -
let us describe it without details. The graph $ \Gamma $ has only one edge, which connects the images of the midpoints 
of the edges of a
Feynman graph with the image of all the vertices; Feynman graphs are drawn not on the base, but on even-sheeted 
base coverings. Note that the edge $ \Gamma $ is not a ribbon.
Since we have inserted the source matrix, instead of vertices, we should consider inflated vertices - 
small disks (stars).

We denote such a graph on the base $ \tig $; unlike the complex matrix model, this graph is not a Feynman diagram.
The first branch point is responsible for pairing the half-rib
(that is, all coverings of degree $ 2d $ have branches of type $ (2^d) $.
The second branch point is responsible for the vertices, the branch profile above this point is given by the Young 
diagram which enters the factor 
$$
\tr\left(C^{\Delta_1}\right)\cdots \tr\left(C^{\Delta_V}\right)=:{\cal C}^{\Delta}
$$
where $ \Delta $ is the branch profile.

The third branch point is responsible for the faces; it is determined by the Young diagram, which is included in
$$
\tr\left((MC)^{\tilde{\Delta}_1}\right)\cdots \tr\left((MC)^{\tilde{\Delta}_F}\right)=:{\cal X}^{\tilde{\Delta}}
$$. 

Combinatorial equation for Feynman diagram of order $2d$:
\be
\alpha\prod_{i=1}^F f_i = \prod_{i=1}^V \sigma_i
\ee
is an equation in the group $ S_{2d} $, and $ \alpha $ has the cycle type $ (2^d) $, each  element 
$ f_i $ is a cycle of length $ \tilde{\Delta} _i $, and each element  $ \sigma_i $ is
a cycle of length $ \Delta_i $. The number of nonisomorphic coverings of $\mathbb{S}^2$ with profiles of type
$ (2^d) $, $ \Delta $ and $ \tilde{\Delta} $ is the Hurwitz number 
$H_{\mathbb{S}^2}\left((2^d),\Delta,\tilde{\Delta}\right)$.
It is natural to assume that
$$
{\cal Z}_N(\bpow,C)=\sum_{d>0} N^{-2nd}
\sum_{\Delta,\tilde{\Delta}\atop |\Delta|=|\tilde{\Delta}|=2d} H_{\mathbb{S}^2}\left((2^d),\Delta,\tilde{\Delta}\right)
{\cal C}^{\Delta}\bpow_{\tilde{\Delta}}
$$
This will be considered in the next paper.

\section*{Acknowledgements}

 Work A.O. was supported by the Russian Science 
Foundation (Grant No.20-12-00195).
The authors are grateful to S. Lando,  
M. Kazarian, D. Vasiliev,
A. Morozov, A. Mironov, L. Chekhov, Yu. Marshall
for helpful discussions. We thank A. Gerasimov, who drew our attention to \cite {Witten},
Yu. Neretin who pointed out \cite{Olshanski-19}.
We are very grateful to G.I.Olshansky who pointed out the works \cite{PerelomovPopov1}, \cite{PerelomovPopov2},
explaining the group-theoretical representation of the relationship (\cite{MM3}).
A.O. is grateful to A.Odzijewicz for the kind hospitality in Bialowieza and
E. Strakhov, who turned his attention to the independent  Ginibre ensembles \cite {Ak1}, \cite {S2}, \cite {S1}.
A.O. thanks Grisha Orlov for children's drawings.

\appendix


\section{Appendices. Definitions and a review of known results \label{definitions}}

When setting out the general information in this Appendix, we basically follow the work of \cite{NO2020}.

\subsection {Hurwitz Numbers.\label{Definition}}

\vspace {2ex}

Hurwitz number is the weighted number of branched coverings of a surface with a prescribed topological
type of critical values.  
Hurwitz numbers of oriented surfaces without boundaries were introduced by Hurwitz
at the end of the 19th century.
Later it turned out that they are closely related to the  module spaces of Riemann
surfaces \cite{ELSV}, to the integrable systems \cite{Okounkov-2000}, modern models of mathematical physics
[matrix models], and closed topological field theories \cite{Dijkgraaf}. In this paper we will consider only the
Hurwitz numbers over compact surfaces
without boundary. The definition and important
properties of Hurwitz numbers over arbitrary compact (possibly with boundary) surfaces were suggested in \cite{AN}.

Clarify the definition. Consider a branched covering $\varphi:P\rightarrow\Omega$ of degree $d$ over a compact surface
without boundary. In the neighborhood of each point $z\in P$, the map $\varphi$ is topologically equivalent to the
complex map $u\mapsto u^p$ in the neighborhood of $u=0\in\mathbb{C}$. The number $p=p(z)$ is called degree of the
covering $\varphi$ at the point $z$. The point $z\in P$ is called \textit{branch point} or  \textit{critical point}
if $p(z)\neq 1$. There are only a finite number of critical points. The images $\varphi(z)$ of any critical point is 
called \textit{critical value}.

Let us associate with a point $s\in\Omega$ all points  $z_1,\dots,z_\ell\in P$ such that $\varphi(z_i)=s$. Let
$p_1, \dots,p_\ell$ be the degrees of the map $\varphi$ at these points. Their sum $d=p_1 +\dots+p_\ell$ is equal
to the degree $d$ of $\varphi$. Thus, to each point $s\in S $ there corresponds a partition $d=p_1 +\dots+p_\ell$ of
the number $d$. By ordering the degrees $ p_1\geq \dots\geq p_\ell >0$ at each point
$s\in\Omega $, we can introduce the Young diagram $\Delta^s=[p_1, \dots, p_\ell]$ of degree $d$ with
$\ell=\ell(\Delta^s)$ number of lines of length $p_1\dots, p_\ell$. The Young diagram $\Delta^s$ is called 
\textit {topological type} of the value $s$.
The value of $s$ is critical if not all $p_i$ are equal to $1$.

Let us note that the Euler characteristics $\e(P)$ and $\e(\Omega)$ of the surfaces $P$ and $\Omega$ are related
by the Riemann-Hurwitz relation:
\be
\e(P)=\e(\Omega)d +\sum\limits_{z\in P} \left(p(z)-1\right).
\ee
or, the same
\be\label{RHur}
\e(P)=\e(\Omega)d +\sum\limits_{i=1}^\fa \left( \ell(\Delta^{s_i})-d\right).
\ee
where $s_1,\dots,s_{\fa}$ are critical values.

An \textit{equivalence} between coverings $\varphi_1:P_1\rightarrow\Omega $ and $\varphi_2: P_2\rightarrow\Omega$ is
called a homeomorphism $F:P_1\rightarrow P_2$ such that $\varphi_1=\varphi_2F$. Coverings are considered
\textit {equivalent}, if there is an equivalence between them. The equivalence of a covering with yourself is called
an \textit{automorphism} of the covering. Automorphisms of the covering $\varphi$ form a group $\texttt{Aut}(\varphi)$
of a finite order $|\texttt{Aut}(\varphi)|$. Equivalent coverings have isomorphic groups of automorphisms.

Fix now points of $s_1, \dots, s_\fa\in\Omega$ and Young diagrams $\Delta^1,\dots,\Delta^\fa$ of degree $d$. Consider the
set $\Phi$ of all equivalence classes of coverings for which $s_1,\dots,s_\fa$ are the set of all critical values, and
$\Delta^1,\dots,\Delta^\fa$ are topological types of these critical values. Further, unless otherwise stated, we consider that the surface $\Omega$ is connected

\textit{Hurwitz number} is the number
\begin{equation}\label{disconH} H_{\e(\Omega)}^d(\Delta^1,\dots,\Delta^\fa)=\sum_{\varphi\in\Phi} 
\frac {1} {|\texttt{Aut} (\varphi)|}.
\end{equation}
It is easy to prove that the Hurwitz number is independent of the positions of the points $s_1, \dots, s_\fa$ 
on $\Omega$.
It depends only on the Young diagrams of $\Delta^1,\dots,\Delta^\fa$ and the Euler characteristic $\e=\e(\Omega)$.
Therefore instead of $H_\Sigma$ we shall write $H_\e$ below, where $\e$ is the Euler characteristic
 of $\Sigma$; in particular we shall write $H_2$ instead of $H_{\mathbb{S}^2}$ and $H_1$ instead of $H_{\mathbb{RP}^2}$.

\vspace {2ex}

\vspace {2ex}

\subsection {Hurwitz numbers and symmetric group.}

Describe now Hurwitz numbers $H_{e}^d(\Delta^1,\dots,\Delta^\fa)$ in terms of the center $Z\mathbb{C}[S_d]$ of the
group algebra $\mathbb{C}[S_d]$ of the symmetric group $S_d$. The action of a permutation $\sigma\in S_d$ on a set
$T$ of $d$ elements splits $T$ into $\ell$ orbits consisting of $\Delta_1,\dots,\Delta_{\ell}$ elements, where
$\Delta_1+\dots+\Delta_{\ell}=d$. The Youn diagram $[\Delta_1,\dots,\Delta_{\ell}]$ we will call
\textit{a cyclic type of $\sigma$}. All permutations of a cyclic type $\Delta$ form a   conjugate class
$C_\Delta\subset S_d$. Denote by $|C_\Delta|$ the number of elements in $C_\Delta$. The sum $\gC_\Delta$ of
elements of the conjugate class $C_{\Delta}$ belongs to the center of the algebra
$Z\mathbb{C}[S_d]$. Moreover, the sums $\gC_\Delta$ generate the vector space $Z\mathbb{C}[S_d]$.

The correspondence $\Delta\leftrightarrow\gC_\Delta$ gives a isomorphism between vector spaces $Y_d$ and
$Z\mathbb{C}[S_d]$. It transfers the structure of algebra to $Y_d$. We will keep it in mind in this section,
speaking about multiplication on $Y_d$.

\vspace {1ex}

Describe now the Hurwitz number $H_{2}^d(\Delta^1,\dots,\Delta^\fa)$ of the sphere $S^2$ in terms of the algebra $
Z\mathbb{C}[S_d]$. Consider different points $\{p_1, \dots, p_\fa \}$ of $S^2$ and $ p\in S^2\setminus\{p_1, 
\dots, p_\fa \}$. Consider the
standard generators of the fundamental group $\pi_1(S^2\setminus\{p_1,\dots, p_\fa \},p)$. They are represented
by simple closed pairwise disjoint contours $\gamma_1,\dots,\gamma_\fa$ with a beginning and an end in $p$, which
bypass the points $p_1,\dots,p_\fa $ and $\gamma_1\dots \gamma_\fa=1$.

Consider now the covering $\varphi:\widetilde{\Omega}\rightarrow S^2$ of the type $(\Delta^1,\dots,\Delta^\fa)$ with
critical values $p_1\dots p_\fa$. The complete preimage of $\varphi^{-1}(p)$ consists of $d$ points $q_1,\dots,q_d$. A
going around the contour $\gamma_i$ get a permutation $\sigma_i\in S_d $ of $q_1,\dots,q_d$ . The conjugacy class of $
\sigma_i$ is described by a Young diagram $\Delta^i$. Moreover, the product $\sigma_1\dots\sigma_\fa$ gives an identical
permutation. Thus, a covering of a sphere of type $(\Delta^1,\dots,\Delta^\fa)$ generates an element of the set
$$
M=M(\Delta^1,\dots,\Delta^\fa)=
\{(\sigma_1,\dots,\sigma_\fa)\in(S_d)^\fa|\sigma_i \in\Delta^i(i = 1,\dots,\fa);\sigma_1\dots\sigma_\fa=1\}.
$$
Moreover, the equivalent coverings generate elements of $M$ that conjugated by some permutation $\sigma\in S_d$.

Construct now the inverse correspondence, from conjugation classes of $M(\Delta^1,\dots,\Delta^\fa)$ to equivalent
classes of coverings $\varphi:\widetilde{\Omega}\rightarrow S^2$ of the type $(\Delta^1,\dots, \Delta^\fa)$ with
critical values $p_1,\dots,p_\fa$.
Cuts $r_i\subset S^2$ between points $p$ and $p_i$ inside the contour $\gamma_i$ generate a cut sphere
$\widehat{S}=S^2\setminus \bigcup\limits_{i=1}^d r_i$.

Correspond now the covering which corresponds to  $(\sigma_1,\dots,\sigma_\fa)\in M$. For this we consider $d$
copies of the cut sphere $\widehat{S}$, number them, and glue its boundaries according to the permutations
$\sigma_1,\dots,\sigma_\fa$. This gives a compact surface $P$. Moreover, the correspondances between the copies of
$\widehat{S}$ and $\widehat{S}$ generate the covering $\varphi:P\rightarrow S^2$, of type $(\Delta^1,\dots,\Delta^\fa)$.
Conjugated by $\sigma\in S_d$ of the set $(\sigma_1,\dots,\sigma_\fa)$ generate equivalent covering.

Thus
$$
H_{\e(S^2)}^d (\Delta^1, \dots, \Delta^\fa) =
\sum_{\varphi\in\Phi(\Delta^1, \dots, \Delta^\fa)} \frac {1} {|\texttt{Aut} (\varphi)|} =
\sum_{(\sigma_1, \dots,\sigma_\fa) \in \widetilde{M}} \frac{1}{| \texttt {Aut} (\sigma_1, \dots,\sigma_\fa)|}.
$$
where $\widetilde{M}$ is the set of conjugated classes of $M$ by $S_d$ and $\texttt{Aut}(\sigma_1,\dots,\sigma_\fa)$
is the stabilizer of $(\sigma_1, \dots,\sigma_\fa)$ by these conjugations.

On the other hand,
$$
\sum_{(\sigma_1, \dots,\sigma_\fa) \in \widetilde{M}} \frac{1}{| \texttt{Aut}
(\sigma_1, \dots,\sigma_\fa)|}=\frac{1}{d!}| M (\sigma_1, \dots,\sigma_\fa) |=<\Delta^1\dots\Delta^\fa>.
$$

Thus,
\begin{equation}\label{sruct.cont} H_{\e(S^2)}^d (\Delta^1, \dots,\Delta^\fa)= <\Delta^1\dots\Delta^\fa>.
\end{equation}

For arbitrary closed connected surface $\Omega$ this relation turns into \cite{AN}
\begin{equation} H_{\e(\Omega)}^d (\Delta^1, \dots,\Delta^\fa)= <\Delta^1\dots\Delta^\fa U^{2-\e(\Omega)}>.
\end{equation}
where
$ U =\sum \limits_ {\sigma \in S_d}\sigma^2 $ \cite{AN, AN2008}.

A proof for arbitrary $\Omega$ is practically the same that for $\Omega=S^2$. It needed only change the relation
$\sigma_1\dots\sigma_\fa=1$ to relations for standard generators in $\pi_1(\Omega,p)$. For orientable $\Omega$ this
is $[a_1,b_1]\dots[a_g,b_g]\sigma_1\dots\sigma_\fa=1$; for non-orientable $\Omega$ this is
$c_1^2\dots c_g^2\sigma_1\dots\sigma_\fa=1$.

In particulary
\begin{equation}\label{HurGen}
H^d_{\e(\mathbb{R}P^2)} (\Delta^1, \dots,\Delta^\fa)
= <\Delta^1\dots\Delta^\fa U>.
\end{equation}

\vspace {2ex}

\subsection{Hurwitz numbers and representation theory \label{Hurwitz numbers and representation theory}}

Formula (\ref{HurGen}) permits to describe Hurwitz numbers in term of the characters of symmetric groups. The
corresponding formula is
\be\label{Mednykh}
H_{\e}^d (\Delta^1, \dots, \Delta^\fa) =
(d!)^{-\e} |C_{\Delta^1}| \dots |C_{\Delta^\fa}|\sum \limits _ {\chi} \frac { \chi (\gC_{\Delta^1})
\dots \chi (\gC_{\Delta^\fa})} {\chi (1)^{\fa-\e}}.
\ee
where summation is carried out over all characters of irreducible representations of the group $S_d$ and $|{C}_\Delta|$  is the cardinality of the set of elements $S_d$ of cyclic type $\Delta$.

The first versions of the formula in the language of symmetric groups appeared in the works of Frobenius and Schur
\cite{Fr,FS}. Geometric
iteration relating to arbitrary surfaces turns appeared in \cite{M1,M2}.
We now give a sketch of the proof of formula (\ref{Mednykh}).

Any partition $\lambda$ of weight $d$ generate a irreducible representation of $S_d$ of dimension $\dim \lambda$.
Let $\chi(\lambda)$ be the character of this representation. Then $\dim\lambda =\chi_\lambda(\gC_{[1,\dots,1]})$.
For any Young diagrams $\Delta$ and $\lambda$, we define the \textit{normalized character}:
\be\label{normalized-characters}
\varphi_\lambda(\Delta) :=| {C}_\Delta | \frac{\chi_\lambda(\Delta)}{\dim\lambda}.
\ee

The known orthogonality relations for the characters are \cite{Mac}
\be\label{orth1}
\sum_{\lambda} \left(\frac{{\rm\dim}\lambda}{d!}\right)^2\varphi_\lambda(\mu)\varphi_\lambda(\Delta) =
 \frac{ \delta_{\Delta,\mu} }{z_{\Delta}}
\ee
and
\be\label{orth2}
\left(\frac{{\rm\dim}\lambda}{d!}\right)^2
\sum_{\Delta} z_\Delta\varphi_\lambda(\Delta)\varphi_\mu(\Delta) =
\delta_{\lambda,\mu}
\ee
where $d=|\Delta|=|\lambda|$ and

\be\label{z_Delta}
z_\Delta=\prod_{i}m_i!i^{m_i} =\frac{d!}{|C_\Delta|}
\ee
is the order of the automorphism group of the Young diagram $\Delta$.
(In this formula $m_i$ is the number of lines of length $i$ in $\Delta$.)

Elements
\be\label{idempotents}
 \gF_\lambda = \left(\frac{{\rm\dim}\lambda}{d!} \right)^2\sum_\Delta z_\Delta\varphi_\lambda(\Delta) \gC_\Delta
\ee
form the basis of idempotent of $Z\mathbb{C}[S_d]$, that is
\be\label{idempotent}
\gF_\lambda \gF_\mu =0,\quad \mu\neq \lambda,\qquad
\gF_\lambda^2=\gF_\lambda
\ee
Further,
\be\label{class-via-idempotents}
 \quad \gC_\Delta = \sum_\lambda  \varphi_\lambda(\Delta)  \gF_\lambda
\ee
and therefore
\be
\gC_{\Delta^1}\cdot \gC_{\Delta^2}=\sum_\lambda \varphi_\lambda(\Delta^1)\varphi_\lambda(\Delta^2)\gF_\lambda
=\sum_\Delta H_2(\Delta^1,\Delta^2,\Delta)z_\Delta\gC_\Delta
\ee
Moreover
\be
< \gC_{\Delta^1}\cdots \gC_{\Delta^\fa} U^{2-\e} > =\sum_\lambda
\varphi_\lambda(\Delta^1)\cdots \varphi_\lambda(\Delta^\fa)< \gF_\lambda U^{2-\e} >
\ee
and
\be
< \gF_\lambda U^{2-\e}> =\left( \frac{{\rm dim}\lambda}{|\lambda|!}\right)^{\e}
\ee

Therefore
\[
H_{\e({\Sigma})}(\Delta^1, \dots, \Delta^\fa) = < \gC_{\Delta^1}\cdots \gC_{\Delta^\fa}>_{{\Sigma}} =
< \gC_{\Delta^1}\cdots \gC_{\Delta^\fa} U^{2-\e({\Sigma})} >
\]
\be\label{Mednykh-Hurwitz}
=\sum_\lambda
\varphi_\lambda(\Delta^1)\cdots \varphi_\lambda(\Delta^\fa)\left( \frac{{\rm dim}\lambda}{|\lambda|!}\right)^{\e}
\ee
that is equivalent to (\ref{Mednykh}).

From (\ref{Mednykh-Hurwitz}) and (\ref{orth1}) we get
$$
 H_2(\Delta^1,\Delta) =\frac{\delta_{\Delta^1,\Delta}}{z_\Delta}
$$.

For the number $D(\Delta)$ describing the M\"{o}bius cut, we get
\be\label{D(Delta)}
D(\Delta)=z_\Delta H_1(\Delta)
\ee
where $H_1(\Delta)$ is the Hurwitz number counting the covering of the real projective plane
$\mathbb{RP}^2$ with one critical value with the ramification profile $\Delta$.

Formulas (\ref{Mednykh-Hurwitz}) and (\ref{orth2}) allow us to give an independent proof of the fact
that Hurwitz numbers satisfy the axioms of Hurwitz
topological field theory:
\noindent
\bp\label{Hurwitz-down-Lemma}
Let us define numbers $H_{\e({\Sigma})}(\Delta^1, \dots, \Delta^\fa)$ by
(\ref{Mednykh-Hurwitz}).
Consider the set of partitions $\Delta^i,\,i=1,\dots,\fa_1+\fa_2$ of the same weight $d$. The nect relation is called 
the handle cut relation
\begin{eqnarray}\label{handle-cut}
H_{\e  -2}(\Delta^{1},\dots,\Delta^{\fa}) = \sum_{\Delta\atop |\Delta|=d}
H_{\e}(\Delta^{1},\dots,\Delta^{\fa},\Delta,\Delta)z_\Delta
\\
= \sum_{\Delta\atop |\Delta|=d}\frac{
H_{\e}(\Delta^{1},\dots,\Delta^{{\fa}},\Delta,\Delta)}
{H_2(\Delta,\Delta)}\,.\nonumber
\end{eqnarray}
\begin{eqnarray}\label{Hurwitz=Hurwirz-Hurwitz}
H_{\e_1+\e_2 -2}(\Delta^{1},\dots,\Delta^{{\fa}_1+{\fa}_2}) = \sum_{\Delta\atop |\Delta|=d}
H_{\e_1}(\Delta^{1},\dots,\Delta^{{\fa_1}},\Delta)z_\Delta
H_{\e_2}(\Delta,\Delta^{{\fa}_1+1},
\dots,\Delta^{{\fa}_1+{\fa}_2})\\
 = \sum_{\Delta\atop |\Delta|=d}\frac{
H_{\e_1}
\left(
\Delta^{1},\dots,\Delta^{\fa_1},\Delta
\right)
H_{\e_2}
\left(
\Delta,\Delta^{\fa_1+1},\dots,\Delta^{\fa_1+\fa_2}
\right)
}
{H_2(\Delta,\Delta)}\,.\nonumber
\end{eqnarray}
\begin{eqnarray}
\label{Hurwitz-down}
H_{\e-1}(\Delta^{1},\dots,\Delta^{{\fa}})=
\sum_{\Delta}\,
H_{\e}(\Delta^{1},\dots,\Delta^{{\fa}},\Delta) D(\Delta)\\
=
\sum_{\Delta}\,\frac{
H_{\e}(\Delta^{1},\dots,\Delta^{{\fa}},\Delta)  H_{1}(\Delta)}{H_2(\Delta,\Delta)}\quad ,\nonumber
\end{eqnarray}
where $\frac {H_{1}(\Delta)}{H_2(\Delta,\Delta)}= D(\Delta) $ 
are rational numbers:
\be\label{delta(Delta)}
D(\Delta)= z_\Delta H_{1}(\Delta)
=\sum_{\lambda \atop |\lambda|=|\Delta|} \chi_\lambda(\gC_{\Delta})
\ee
see (\ref{Mednykh}).
\ep

\noindent

\subsection{On  Moebius strip and on handle insertitions \label{Mobius'}}

We have
\be\label{vac-tau-BKP'}
\frac{\det^{1/2}\dfrac{1+X}{1-X} }{\det^{1/2}\left( I_N \otimes I_N - X\otimes X\right)}=
 \sum_\lambda \,s_\lambda(X)=
 e^{\frac 12 \sum_{m>0} \frac 1m\left(\tr X^m\right)^2 + \sum_{m>0,{\rm odd}}\frac{1}{m}\tr X^m}
 =\sum_{\Delta\in{\cal P}} \bpow_{\Delta} D(\Delta),
\ee
where ${\cal P} $ is the  set of all partitions,
where we denote $p_m=\tr(X^m)$ and $\bpow_\Delta=p_{\Delta_1}p_{\Delta_2}\cdots $. 
The function defined in (\ref{vac-tau-BKP'}) written in $\{p\}$ variables\footnote{It was written down
in \cite{OST-I} as the simplest nontrivial example of the BKP hypergeometric tau function.}
was used in \cite{NO-LMP} as the generation function for 1-point Hurwitz numbers for $\mathbb{RP}^2$.

Let us write down the generating function of 1-point Hurwitz number for $\Sigma=\mathbb{RP}^2$.
First, let us write down the simplest case of a single branch point related to all $r=1$ and $N=\infty$.
This case is generated by $\tau_1^{\rm B}$, where it is reasonable to produce the change $p_m\to h^{-1}c^m p_m$.
We get
\be\label{single-branch-point}
e^{\frac {1}{h^2}\sum_{m>0} \frac {1}{2m}p_m^2 c^{2m} +\frac 1h\sum_{m {\rm odd}} \frac 1m p_m c^m}=
\sum_{d>0} c^d 
\sum_{\Delta\atop |\Delta|=d} h^{-\ell(\Delta)} \bpow_\Delta
H^{1,a}(d;\Delta)\,,
\ee
where $a=0$ if $\Delta=(1^d)$, and $a=1$ otherwise. Then $H^{1,1}(d;\Delta)$ is the Hurwitz number 
describing a $d$-fold covering of $\mathbb{RP}^2$ with a single
branch point of type $\Delta=(d_1,\dots,d_l)$, $|\Delta|=d$ by a (not necessarily connected) Klein surface of
Euler characteristic $\textsc{e}'=\ell(\Delta)$. For instance, for $d=3$, $\textsc{e}'=1$, we get
$H^{1,1}(3;\Delta)=\delta_{\Delta,(3)}/3$.
For unbranched coverings $a=0$, $\textsc{e}'=d$.

Next note that the exponent on the left-hand side may be rewritten as the generating series of the
connected Hurwitz numbers
\[
 \frac{1}{h^2}\sum_{d=2m}c^{2m} p_m^2 H_{\rm con}^{1,1}\left(d;(m,m)\right) +
 \frac{1}{h} \sum_{d=2m-1} c^{2m-1}p_{2m-1} H_{\rm con}^{1,1}\left(d;(2m-1)\right)\,,
\]
where $H_{{\rm con}}^{1,1}$ describes a $d$-fold covering either by the Riemann
sphere ($d=2m$) or by the projective plane ($d=2m-1$). These are the only ways to cover $\mathbb{RP}^2$
by a connected surface for the case of a single branch point.
The geometrical meaning of the exponent in (\ref{single-branch-point}) may be explained as follows. The projective plain may be viewed as the unit disk with the identification
of the opposite points $z$ and $-z$ on the boundary $|z|=1$. If we cover the Riemann sphere by the Riemann sphere $z\to z^m$, we get
two critical points with the same profiles. However, if we cover $\mathbb{RP}^2$ by the Riemann sphere, then we have the composition of the
mapping $z\to z^{m}$ on the
Riemann sphere and the factorization by antipodal involution $z\to - {1}/{\bar z}$. Thus we have the ramification profile $(m,m)$
at the single critical point $0$ of $\mathbb{RP}^2$.
The automorphism group is the dihedral group of order $2m$, which consists of rotations by ${2\pi }/{m}$ and antipodal involution
$z\to -{1}/{\bar z}$.
Thus we get that $H_{\rm con}^{1,1}\left(d;(m,m)\right)={1}/{2m}$, which is the factor in the first sum in the exponent in
(\ref{single-branch-point}). Now let us cover $\mathbb{RP}^2$ by $\mathbb{RP}^2$ via $z\to z^d$. For even $d$, we have the critical point
$0$, and in addition each point of the unit
circle $|z|=1$ is critical (a folding), while from the beginning we restrict our consideration to isolated critical points.
For odd $d=2m-1$, there is
a single critical point $0$, the automorphism group consists of rotations through the angle ${2\pi}/(2m-1)$. Thus in this case
$H^{1,1}\left(d;(2m-1)\right)={1}/(2m-1)$, which is the factor in the second sum in the exponent in (\ref{single-branch-point}).

\subsection {The generating function for simple Hurwitz numbers.}
\vspace {2ex}
Important applications of Hurwitz numbers are associated with the corresponding generating functions for 1- and 2-
Hurwitz numbers. A (disconnected) simple 1-Hurwitz number $h_ {m, \Delta}^{\circ}$ is  Hurwitz number
$ H_{\mathbb{S}^2} (\Delta, \g_1, \dots, \g_m) $, where $\g_1=\dots=\g_m=[2,1, \dots, 1]$
(not to be confused with the designation of the graph in the main text)
and where $|\Delta|=\cdots =|\Gamma_m|=d$.

The generating function for 1-Hurwitz numbers depends on an infinite number of formal variables $ p_1, p_2, \dots $.
We associate the Young diagram with $ \Delta $ with strings of length $ d_1, \dots, d_k $ with monomial 
$ p_{\Delta} = p_ {d_1}, \dots, p_ {d_k} $. The generating function for 1-Hurwitz numbers is defined as
$$ 
F^{\circ} (u | p_1, p_2, \dots) =\sum \limits_{m = 0}^\infty
\sum \limits_{\Delta}^\infty \frac {u^m} {m!} h_ {m, \Delta}^{\circ} p _ {\Delta}. 
$$
This feature has a number of remarkable properties discovered relatively recently. The first is the relationship
between the $ u $ variable and the $ p_i $ variables.
\begin {equation}\label{K-a-J}
\frac{\partial F^{\circ}} {\partial u} = L^{\circ}F^{\circ},
\end{equation}
where
\be\label{Hopf}
L^{\circ} = \frac {1} {2}\sum \limits_ {a, b = 1}^\infty \Big ((a + b) p_ap_b
\frac {\partial} {\partial p_ {a + b} } + ab p_ {a + b} \frac {\partial^2} {\partial p_a \partial p_b} \Big).
\ee
This relationship was first found in \cite{GJ} by purely combinatorial methods.
It can be also obtained with the help of vertex operators
\cite{PogrebkovSushko} and tau function \cite{JM},
\cite{MM1}, \cite{OS-2000}.
But it also has a geometric explanation \cite{MM3}.
Consider the covering $\varphi:{\Sigma}\rightarrow S^2$ of the type
$(\Delta, \g_1, \dots, \g_m)$. Let $q, p\in S^2$ be the critical points of the covering $\varphi$ corresponding 
to the Young diagrams $\Delta$ and $\g_m$, respectively. Connect the points $q$ and $p$ with a line $l$ without 
self-intersections. The preimage of $\varphi^{-1}(l)$ consists of $d-1$ connected components, exactly one of which 
$\tilde{l}$ contains the critical point $\tilde {p}$ with the critical value $p$. The ends of the component $
\tilde {l}$ are the pre-images of $\tilde {q}_1$ and $\tilde {q}_2$ points of $q$.

We will now move the point $p$ along the line $l$ in the direction of the point $q$, continuously changing the 
covering of $\varphi$ accordingly. As a result, we get a covering $ \varphi '$ of the type
$ (\Delta', \g_1, \dots, \g_ {m-1}) $. Let us see what kind of Young diagram $ \Delta '$ can do this.
Let $ \tilde {q}_1 = \tilde {q}_2 $ and $ c $ be the branching order of the covering $ \varphi $ at this point $
\tilde {q} = \tilde {q}_1 = \tilde { q}_2 $. In the process of deformation of the covering of $ \varphi $ into the
covering of $ \varphi '$, orders other than $ \tilde {q} $ of critical points will not change. The point
$ \tilde {q} $, as a result of the deformation, splits into 2 points with branching orders $ a $ and $ b $,
where $ a + b = c $. Thus, the monomial $ p_ \Delta $ becomes a monomial
$ p_ap_b \frac {\partial p_ \Delta} {\partial p_c} $.

Suppose that the critical points $ \tilde {q}_1 $ and $ \tilde {q}_2 $ do not coincide and the orders of their
branching are $ a $ and $ b $, respectively. Then, as before, in the process of deformation the covering of
$ \varphi $ into a covering of $ \varphi '$, the orders of critical points other than $ \tilde {q}_1 $ and $
\tilde {q}_2 $ will not change. The points $ \tilde {q}_1 $ and $ \tilde {q}_2 $ as a result of the deformation
will be transferred to one critical point of order $ c = a + b $. Thus, the monomial $ p_ \Delta $ becomes a monomial
$ p_c\frac {\partial^2 p_ \Delta} {\partial p_ap_b}$.
Summation over all possible equivalence classes of covers of all types of $ (\Delta, \g_1, \dots, \g_m) $
and all their deformations into covers of the types $ (\Delta ', \g_1, \dots, \g_ {m-1} ) $ just gives the relation
(\ref{K-a-J}).

Differential properties of the function $F^{\circ} (u | p_1, p_2, \dots)$ were investigated in 
\cite{MM1,MM3,MM4,MM5,AMMN-2011,AMMN-2014}.

\subsection{From cycle-products of the faces to cycle-product of the stars \label{Dual monodromies} }
Here we follow the work \cite{NO-2019}, which describes the transition from the set of monodromies
 (\ref{star-monodromy}) to the set of (dual) monodromies (\ref{face-monodromy}).

 In each face monodromy matrix (\ref{face-cycles-monodromies}), we equate each matrix from 
 the set $ \{Z \} $ to the identity matrix $ N \times N $,
and for the resulting matrices, let's introduce the notation:
\be\label{face-monodromy'}
W_i :=M_i|_{Z_i\to \mathbb{I}_N,\,i=1,\dots,2n}={\cal D}\left[ f_i \right],
\ee
where $ f_a $ is the face cycle with number $ a $. We will call such monodromies words. Like monodromies,
words are defined up to a cyclic permutation. Consider $ W_1 \otimes W_2 \otimes \cdots \otimes W_\fa $
(the order in this tensor product is not important) and the set of involutions $ T_i, \, i = 1, \ dots, n $, which act on this tensor product as follows.
Each involution of $ T_i $ does not affect those $ W_a $ that contain neither $ C_i $ nor $ C_{- i} $. Are possible
two situations. (I) The matrices $ C_i $ and $ C_{- i} $ are in the same word, say, the word $ W_a $. So how can we
to rearrange the matrices with the word cyclically, we bring it to the form $ C_i X C_{- i} Y $, where $ X $ and 
$ Y $ are some matrices. (II) The matrices $ C_i $ and $ C_{- i} $ are included in different words, in this case 
we will write these two words as $ C_iX $ and $ C_{- i} Y $.
Then
\bea
T_i\left[ \cdots \otimes C_i X C_{-i}Y \otimes \cdots \right]=\cdots \otimes C_iX \otimes C_i Y \otimes \cdots
\\
T_i\left[ \cdots  \otimes C_iX \otimes C_i Y \otimes \cdots \right] =\cdots \otimes C_i X C_{-i}Y \otimes \cdots 
\eea
It is easy to see that involutions commute: $ T_i [T_j [*]] = T_j [T_i [*]] $.

If we recall the graph $ \Gamma $ with ribbon edges, then the operation $ T_i $ is as follows. 
It is necessary to ``expand '' tape number $ i $. 
We will assume that this ribbon turned into a rectangle with vertices 1,2,3,4 and the sides of
the ribbon are the arrow $1\longrightarrow 2$ and the arrow $3\longrightarrow 4$. 
As a result of applying of $T_i$ is as follows:
the rectangle 1,2,3,4 has become a new ribbon, but the sides of this ribbon are now paired arrows
$1\longrightarrow 4$ and $3\longrightarrow 2$.

The transformation 
$$
M_1,\dots, M_\fa\, \leftrightarrow \, W_1^*,\dots, W_\V^*
$$
can be obtained purely algebraically in $n$ steps.

Proposition:
\bea\label{cut-or-join}
\prod_{i=1}^n T_i\left[ W_1\otimes W_2 \otimes \cdots \otimes W_\fa \right]=
W_1^*\otimes \cdots \otimes W_\V^*
\\
\prod_{i=1}^n T_i\left[ W_1^*\otimes \cdots \otimes W_\V^* \right]=W_1\otimes W_2 \otimes \cdots \otimes W_\fa
\eea
 This is a manifestation of the equation (\ref{combinatorial-graph}) in the language of dressed-up cycles.
An involution without fixed points $ \prod_{i = 1}^n T_i $ takes the graph $ \Gamma $ to the graph dual to it.

\end{document}